\documentclass[pre,aps,twocolumn]{revtex4-2}
\usepackage{amsfonts}
\usepackage{times}
\usepackage[normalem]{ulem}
\usepackage[
  breaklinks=true,
  colorlinks=false
]{hyperref}
\usepackage{amsmath, amsthm}
\usepackage[final]{graphicx}
\usepackage{amssymb, algorithm, algpseudocode}
\algnewcommand\algorithmicforeach{\textbf{for each}}
\algdef{S}[FOR]{ForEach}[1]{\algorithmicforeach\ #1\ \algorithmicdo}
\usepackage[title]{appendix}
\usepackage{xcolor,comment}

\renewcommand{\Re}{\mathop{\mathrm{Re}}}

\begin{document}

\renewcommand{\UrlFont}{\normalfont}
\title{Dam breaks in the discrete nonlinear Schrödinger equation}

\author{Shrohan Mohapatra}
\affiliation{Department of Mathematics and Statistics, University
of Massachusetts, Amherst, Massachusetts 01003-4515, USA}

\affiliation{Department of Physics, University
of Massachusetts, Amherst, Massachusetts 01003, USA}

\author{Su Yang}
\affiliation{Department of Mathematics and Statistics, University
of Massachusetts, Amherst, Massachusetts 01003-4515, USA}

\author{Sathyanarayanan Chandramouli}
\affiliation{Department of Mathematics and Statistics, University
of Massachusetts, Amherst, Massachusetts 01003-4515, USA}

\author{P. G. Kevrekidis}
\affiliation{Department of Mathematics and Statistics, University
of Massachusetts, Amherst, Massachusetts 01003-4515, USA}

\affiliation{Department of Physics, University
of Massachusetts, Amherst, Massachusetts 01003, USA}

\date{\today}

\begin{abstract}
In the present work we study the nucleation of Dispersive shock waves (DSW) in the {defocusing}, discrete nonlinear Schr{\"o}dinger equation (DNLS), a model of
wide relevance to nonlinear optics
and atomic condensates.
Here, we study the dynamics of so-called dam break problems with step-initial data characterized by two-parameters, one of which corresponds to the lattice spacing, while the other being the right hydrodynamic background.
Our analysis bridges the 
anti-continuum limit of vanishing 
coupling strength with the well-established
continuum integrable one.
To shed light on the transition  between the extreme limits, we theoretically deploy Whitham modulation theory, various quasi-continuum asymptotic reductions of the DNLS and 
existence and stability analysis and 
connect our findings with systematic 
numerical computations. Our work unveils a sharp threshold in the discretization across which qualitatively continuum  dynamics from the dam breaks are observed. Furthermore, we observe a rich multitude of wave patterns in the small coupling limit including unsteady (and stationary) Whitham shocks, traveling DSWs,  discrete NLS kinks and dark 
solitary waves, among others. Besides, we uncover the phenomena of DSW breakdown and the subsequent formation of multi-phase wavetrains, due to generalized modulational instability of \textit{two-phase} wavetrains. We envision
this work as a starting point towards
a deeper dive into the apparently rich DSW phenomenology in a wide
class of DNLS models across different 
dimensions and for different nonlinearities.
\end{abstract}

\maketitle

\section{Introduction}
The discrete nonlinear Schr\"{o}dinger (DNLS) equation~\cite{kev09} has been a prototypical model
for the study of nonlinear phenomena
involving the interplay of lattice diffraction
and nonlinearity in optics and Bose-Einstein condensation (c.f. \cite{lederer_discrete_2008,morsch_dynamics_2006}).
In the former, the self-trapping of light through the {Kerr}-effect is well known.  
Accordingly, in large arrays of weakly coupled waveguides, 
where DNLS models the slowly varying envelope  of an optical 
beam,
\textit{discrete} solitary wave structures 
were observed in~\cite{eisenberg1998discrete}. 
Similarly, in atomic Bose-Einstein condensates (BECs),
the presence of optical lattice potentials and
of a mean-field-induced cubic nonlinearity led to 
solitary wave dynamics that continue to be explored 
experimentally to this
day~\cite{elmar}. 
Furthermore, the DNLS could also be viewed as a discrete approximation to continuum NLS dynamics, given a sufficiently small lattice spacing~\cite{AblowitzPrinariTrubatch}. 
 
Another front of important recent developments in 
dispersive hydrodynamics has concerned the
exploration of dispersive shock waves (DSW)~\cite{hoefer_dispersive_2009,el_dispersive_2016-1}.
The latter are universal excitations emerging from competing nonlinear and dispersive wave effects.
They occur across various disciplines including fluid mechanics~\cite{trillo_observation_2016,maiden_observation_2016,el_transformation_2012,Ablowitz2011,ablowitz_nonlinear_2012,fibich2015nonlinear}, optical beam propagation ~\cite{wan_dispersive_2007,bendahmane_piston_2022,fatome_observation_2014}, nonlinear waves in cold atomic platforms~ \cite{hoefer_dispersive_2006,chang_formation_2008,hoefer_matter-wave_2009,kevrekidis_emergent_2008}, and in the materials science
topic of granular crystals~\cite{molinari_stationary_2009,yasuda_emergence_2017,chong_dispersive_2018} and related Fermi-Pasta-Ulam-Tsingou lattices~\cite{li2021observation}. The study of dispersive shock waves (DSW) in the one-dimensional continuum NLS (and variants) has been the focus of intense research efforts in the past decades \cite{el_dispersive_2016-1,el1995decay,hoefer_shock_2014,el_resolution_2005}.  In particular, the generation of DSW has been investigated by examining the dynamics of so-called {Riemann problems} \cite{el_dispersive_2016-1,leveque_finite_2002}, which correspond to time-evolved states for step-like initial data. As a model encompassing bi-directional dispersive hydrodynamics (DH), the continuum NLS Riemann problem generically emits two families of counter-propagating nonlinear waves \cite{el1995decay}. {Dam breaks} correspond to specific classes of such Riemann problems, whose nomenclature derives from classical hydrodynamics \cite{leveque_finite_2002}. Here, analogized homogeneous ``fluid" states (at rest) separated by a diaphragm initially are set in motion by the sudden rupture of the diaphragm leading to a hydrodynamic-like pressure-gradient driven flow. Dam breaks in the continuum Schr{\"o}dinger equation across non-zero hydrodynamic backgrounds lead to the nucleation of counterpropagating waves in the form of rarefactions and DSWs with an expanding homogeneous background between them, whose formation facilitates the equilibration of the hydrodynamic pressure across the diaphragm \cite{el_dispersive_2016-1}.

In the present work, our aim is to synthesize these
important topics, namely to explore DSWs in the 
DNLS setting, building on important earlier work 
on the subject.
DSW in the discrete defocusing DNLS have  been explored 
not only in the DNLS but also in the broader context of the Salerno equation (a homotopic continuation between the Ablowitz-Ladik and the DNLS models~\cite{salerno1992quantum}), where far-from-continuum effects were observed in distinct parametric regimes defined by a homotopy parameter and wave amplitude, for unit lattice spacing \cite{kamchatnov_dissipationless_2004,konotop1997dark,konotop2000shock}. Such effects included the emission of discrete DSW of either {polarity}, as well as \textit{compound waves}~\cite{kamchatnov_dissipationless_2004-1} among others. 
In some of these early studies, reductions to 
continuum models such as the Korteweg-de Vries 
equation were often guiding the theoretical
analysis~\cite{konotop1997dark,konotop2000shock}.
A more recent work of \cite{panayotaros2016shelf} explores the far-from-continuum dynamics of the discrete NLS, in which transitions to continuum behavior have been tracked (by reducing the lattice spacing appropriately). In particular, the vacuum dam break case revealed the formation of kinks. Such stationary states gradually transition to rarefaction waves for finer spatial discretizations, with less straightforward
to interpret and rather complex wave patterns emerging for intermediate values of the lattice spacing.

In this work, we return to this program aiming
at a systematic categorization and quantification of
its shock wave features. In particular, we wish to
provide a more systematic
catalogue of the possible behaviors, varying 
canonical parameters of the model
including, e.g., the lattice spacing and (effectively)
the height of the jump in the Riemann initial data.
We provide a phenomenological understanding
(and wherever possible a quantitative characterization)
of the regime of small coupling $\beta/u_{-}^2\ll 1$, and furthermore, an understanding of how we exit this extremely rich regime as  $\beta$ is increased towards a domain governed by continuum-NLS like dynamics. In the latter,
integrability practically only allows for relatively
simpler structures such as rarefactions and DSWs. 
We leverage a wide range of tools from Whitham modulation theory  \cite{sprenger2024whitham,hoefer_shock_2014} and quasi-continuum reductions in weakly nonlinear settings,
to existence and stability of specific coherent structures
that spontaneously arise, such as kinks and dark solitary
waves. We corroborate our findings with extensive 
numerical simulations (and accompanying computations
where needed) of the different regimes. We hope for
this work to catalyze and springboard the apparently
deeply rich exploration of dispersive hydrodynamic
phenomena in such canonical lattice nonlinear 
dynamical system settings.

Our presentation is structured as follows. 
In Section~\ref{Section-II}, we provide the setup
and background of the problem. Sections III-V
illustrate some of the theoretical tools that we
bring to bear towards the details of our study. 
Section VI provides the core of our numerical
results and findings in conjunction with our 
theoretical analysis thereof. Finally, section VII
provides a summary of our findings and paves the
way towards a number of future directions for further
study.

\section{Setup and Riemann initial condition}
\label{Section-II}
The  defocusing discrete nonlinear Schr\"{o}dinger (DNLS) equation is given by
\begin{equation} \label{EquationOfMotionForDNLS}
    i \dot{u}_n = - \beta(u_{n+1} + u_{n-1} - 2 u_n) + |u_n|^2 u_n;
\end{equation}
where the factor $\beta = \frac{1}{h^2}$ describes the degree of discreteness of the lattice. More concretely, $\beta \to 0$ would refer to the anti-continuum limit and $\beta >> 1, h << 1$ would correspond to the continuum limit, the latter being a higher-fidelity representation of the continuum NLS dynamics. 
It is worthwhile to note in passing that the discrete model can also be interpolated by the band-limited full-dispersion continuum model \cite{sprenger2024whitham}
\begin{equation}
\label{FDNLS-continuum}
    i\psi_t-\tilde{\omega}(-i\partial_x)\psi-|\psi|^2\psi=0,
\end{equation}
where $u_n(t)=\psi(nh,t)$. Moreover, the effect of the pseudo-differential operator is defined in Fourier space by the frequency ${\omega}(\kappa; h)=2(1-\cos(\kappa h))/h^2$, which possesses a cut-off value at $\omega_{c}=4/h^2$. The study of such variants of
the DNLS is of interest in its own right, although it will not
be the focus of our present work.


Our starting point herein will be the examination of the dynamics of a two-parameter family of dam break problems
\begin{equation} \label{InitialConditionForDNLS}
    u_n(0) = \begin{cases}
        u_- & n < \frac{N}{2} \\
        u_{+} & n \geq \frac{N}{2}
    \end{cases},
\end{equation}
with $u_-> u_+$, to Eq.~\eqref{EquationOfMotionForDNLS}, for each $\beta$.\\
We mention here two relevant remarks regarding the parameters
and boundary conditions of the considered problem:
\begin{itemize}
    \item Given that $u_n(t)$ is an evolutionary solution, we have that $u_n=\sqrt{\beta} U_n(\beta t)$ satisfies the lattice equation
   { \begin{align}
        &iU_n^{'}(\chi)+\\\nonumber &(U_{n+1}-2 U_n +U_{n-1})-|U_n|^2U_n=0,
    \end{align}}
    with unit coupling strength and $\chi=\beta t$. Furthermore, the family of Riemann initial conditions is characterized by only two parameters: the left and right states $U_-$ and $U_+$ respectively.
    Thus, we see that there are only two free parameters \textit{effectively} for the family of Riemann problems to Eq.~\eqref{EquationOfMotionForDNLS}: we choose to fix $u_-=1$, while varying $u_+$ and $\beta$.    \item Due to practical issues associated with computation over an infinite lattice, we pick a sufficiently large lattice and enforce homogeneous Neumann boundary conditions (BCs) $ u_{0}(t) = u_{1}(t), u_{N}(t) = u_{N+1}(t)$ at the ends of the lattice. Practically, the
    boundaries will be sufficiently far that they will not
    play any relevant role in the considerations that follow.
\end{itemize}
Eq.~\eqref{EquationOfMotionForDNLS} subject to homogeneous Neumann BC's possesses two conserved quantities given by the mass (conservation of $L^2$ norm $M_{\text{DNLS}, \{u_n\}} = \sum_{k=1}^{N} |u_k|^2$) and Hamiltonian in the system

\begin{multline} \label{EquationReferenceForHDNLS}
   H_{\text{DNLS}} =  -\sum_{k=1}^{N-1} \beta \operatorname{Re}\{u_k u_{k+1}^*\} + \frac{\beta}{2} |u_1|^2 + \sum_{k=2}^{N-1} \beta |u_k|^2 \\ + \frac{\beta}{2} |u_N|^2 + \frac{1}{4} \sum_{k=1}^{N} |u_k|^4.
\end{multline}

Note the discrete NLS equation possesses a two-parameter family of stable, homogeneous states (plane wave solutions or \textit{one-phase} Stokes waves) of the {form} {$u_n^{(0)}=A\exp(i\Phi)=A\exp\left(i\kappa nh-i(\omega(\kappa; h)+A^2)t\right)$}, where $|\kappa h|\leq \pi/2$ \cite{kivshar1992modulational,sprenger2024whitham}.
A canonical problem corresponds to studying the propagation of  disturbances on these plane waves; which we at first do without making any assumptions on amplitude
\begin{equation}
    u_n =(A+a(nh,t))\exp(i\kappa n h-i\omega(\kappa; h)t-iA^2 t-i\phi(nh,t)),
    \label{pert}
\end{equation}
{where the lattice spacing $h$ sets an intrinsic spatial scale, and $x_n\equiv nh$.}
This problem of setting
up the equations for
$(a,\phi)$ 
was explored in the seminal
work of~\cite{kivshar1992modulational} and its main 
findings will recur in our considerations below.
The Stokes-waves modulation equations (analogous to the hydrodynamic limit in continuum NLS models) can be derived; this was 
considered recently in the work of~\cite{sprenger2024whitham}.
The resulting coupled one-phase modulation system was shown to admit a Riemann invariant form given by  
 \begin{equation} 
 \label{One-phase-modn}r^{(i)}_T+c^{(i)}r^{(i)}_X=0,
 \end{equation}
 with the Riemann invariants $r^{(1,2)}=(2\sqrt{2}/h) E(\frac{\kappa h}{2},2)\mp 2|{{a}}|$; where $X=\delta x$, $T=\delta T$ represent slow space-time scales (with the small parameter $\delta\ll 1$). $E(\cdot,m)$ represents the elliptic integral of the second kind, where $m$ is the elliptic modulus. The hyperbolic speeds are $c^{(1,2)}=2(\sin(\kappa h))/h\mp |{a}|\sqrt{2\cos(\kappa h)}$. Note that in the continuum limit ($h\rightarrow 0$), these become $c^{(1,2)}=2\kappa\mp \sqrt{2}|{a}|$. The Riemann invariants are an important ingredient in the study of Riemann problems in both hyperbolic \cite{leveque_finite_2002} and dispersive-hyperbolic systems~\cite{el_dispersive_2016-1,el_theory_2007}.
 
 {Notably, the DNLS Riemann invariants and hyperbolic speeds turn complex-valued outside the admissible range (which in the
 defocusing case corresponds to $|\kappa h|\leq\pi/2$}) \cite{hays_macroscopic_1994}. This occurrence is  indicative of modulational instability~\cite{kivshar1992modulational} of plane wave states. 
 This, in turn, can be shown to impose restrictions for the domain of existence of simple waves (rarefaction waves) \cite{el_dispersive_2016-1} and also classical dispersive shock waves (governed by continuum NLS DSW jump conditions \cite{hoefer_shock_2014}). The domain (for $u_-=1$) in which \textit{only} continuum NLS-like dynamics emerge from Riemann problems is suggested by the marginal vacuum dam break problem,  where $u_+=0$ in Eq.~\eqref{InitialConditionForDNLS}. According to classical simple wave theory, this (initial) step is resolved through a \textit{single} rarefaction wave across which $r^{(2)}$ is constant (with a local speed of propagation $c^{(1)}$). Imposing the constancy of the second Riemann invariant we have
 \begin{equation}
     (\sqrt{2}/h)E\left(\frac{\kappa h}{2},2\right)=|u_-|=1.
 \end{equation}
 This equation can be solved provided $|\kappa h|\leq \pi/2$, corresponding to the regime of \textit{hyperbolicity} of the modulation equations Eqs.~\eqref{One-phase-modn} \cite{hays_macroscopic_1994}. Thus, the suggested edge of existence of rarefaction waves
 is marked by the limiting case $\kappa h=\pi/2$, when one obtains the constraint on the lattice spacing
 $ h\leq  \frac{\sqrt{2}E(\pi/4,2)}{|u_-|}\approx {0.8472}{}$ (i.e. $\beta \geq \frac{1}{2 E^2(\pi/4,2)}\approx 1.4$). 
 This is captured, e.g., in the top two panels of
 Fig.~\ref{fig:DNLS_out_of_continuum}.
 {Outside this interval, numerical simulation points to a plethora of new features that emerge which have no continuum NLS analog (see Fig.~\ref{fig:DNLS_out_of_continuum}), whose characterization will form the focus of this paper. We refer to $\beta_c=\frac{1}{2 E^2(\pi/4,2)}$ as the small-to-large coupling threshold. To corroborate this analytically derived threshold, we show numerical simulations above and below this threshold in Fig.~\ref{fig:DNLS_out_of_continuum}. 
 The simple wave profile can be obtained from the implicit equation {for the self-similar one-phase modulation solution}
 \begin{equation}
 c^{(1)}\left(r^{(1)}(x/t),r^{(2)}_{1}\right)=(x/t), 
\end{equation}
{where $c^
 {(1)}(r^{(1)},r^{(2)})$ is the \textit{slow} hyperbolic velocity, and  $r^{(2)}_{1}$ is the value of the second Riemann invariant evaluated at the left hydrodynamic state, where $u_-=1, \kappa_1=0$.}
 Clearly, above the threshold, continuum-NLS like dynamics are observed (c.f. \cite{el1995decay} for a cataloguing of continuum NLS dynamics). On the other hand,  below the threshold, say for $\beta=1$, for $u_+=0$ (in Fig.~\ref{fig:DNLS_out_of_continuum}(C)), a qualitatively distinct traveling-wave feature is seen to propagate downstream. Moreover, for $u_+=0.5$ higher-order dispersive effects begin to emerge (as seen in panel D) which distort the rarefaction and DSW profiles.  
 \begin{figure}
     \centering
     \includegraphics[width=\linewidth]{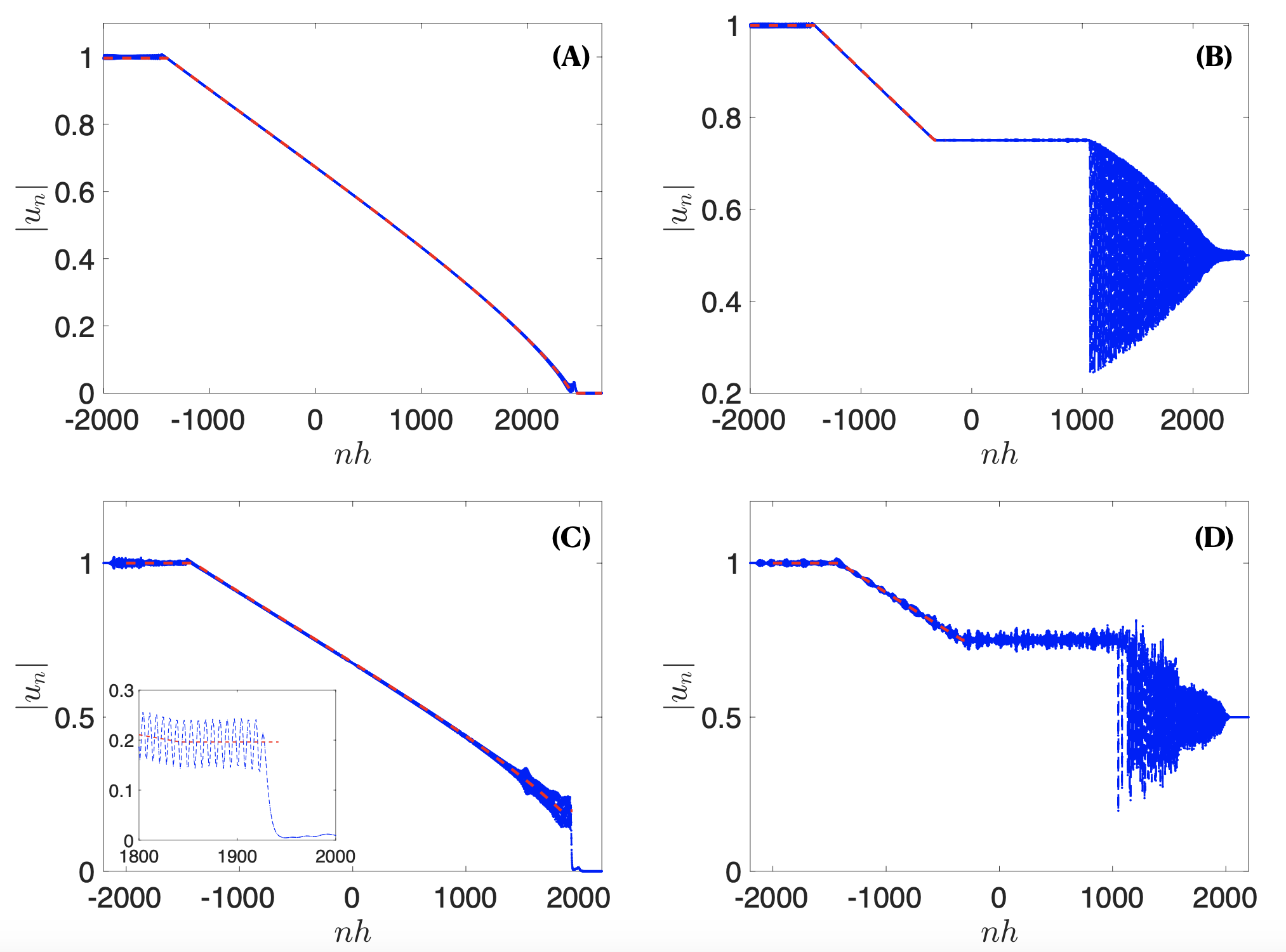}
     \caption{For $\beta=1.5>1/(2E^2(\pi/4,2))$ (Small-to-large coupling threshold), we observe continuum NLS-like-wave
     patterns, as shown in the panels (A) a rarefaction wave emanates from the dam break with $u_+=0$ and (B) a (left) right propagating (rarefaction) DSW emerges for $u_+=0.5$ respectively. Below the small-to-large coupling threshold (i.e., 
     for $\beta=1$), we show snapshots of DNLS simulations. In the vacuum case ($u_+=0$), (C) besides the propagating rarefaction wave, a novel, right propagating traveling wave feature is seen to emerge (zoomed view in the inset). On the other hand, when $u_+=0.5$, (D) higher-order dispersive effects are witnessed which alter the (left) right propagating (rarefaction) DSW patterns. In all the panels, the DNLS simulations at $t=1000$ are shown in the blue dashed curve, with the $r^{(1)}$-simple wave of the one-phase modulation system overlaid in a red dashed line. }
     \label{fig:DNLS_out_of_continuum}
 \end{figure}
 

\section{Linear dispersion relation of two-phase waves} 
Wave propagation in the context of Schrödinger systems is characterized by two distinct phases; the first corresponding to $\Phi=A \exp(i\kappa nh -i(\omega+A^2) t)$ (the homogeneous backgrounds/Stokes waves), while the second phase is associated with the amplitude $a(x_n,t)$ (or phase $\phi(x_n,t)$)
of Eq.~(\ref{pert}) (Eq.~\eqref{FDNLS-continuum}).
Having looked at the one-phase modulation system in Sec.~\ref{Section-II} (Eqs.~\eqref{One-phase-modn}), we next shed light on some characteristics of the dispersion relation of the second phase.
The linear dispersion relation to this system can be sought by substituting $a=a_0 \exp(i(knh-\Omega t))$ and $\phi=\phi_0 \exp(i(knh-\Omega t))$
in the equations resulting from Eq.~(\ref{pert}) (derived,
e.g., in~\cite{kivshar1992modulational}), 
where $k\in [-\pi/h,\pi/h]$ (and $knh-\Omega t$ is the second phase of two-phase periodic waves), which then yields (upon eliminating $\phi_0,a_0$)
$n$
\begin{align}
\label{Disp-relation-band-limited}
    &\Omega_{h,\pm} = 2\beta \sin(kh)\sin(\kappa h)\\\nonumber &\pm 4\sqrt{\beta}\sin\left(\frac{kh}{2}\right)\sqrt{\cos(\kappa h)}\sqrt{\beta\sin^2\left(\frac{kh}{2}\right)\cos(\kappa h)+\frac{A^2}{2}},
\end{align}
with $K=kh$. {In the long wave limit $K \ll 1$, one can expand the (band-limited) dispersion relation} 
\begin{align}
\label{Long-wave-exp}
    \Omega_h &\sim \tilde{\alpha}_1 k+\tilde{\alpha}_3 k^3+\tilde{\alpha}_5 k^5+{o}(k^5),
\end{align}
with coefficients 
\begin{equation}
    \tilde{\alpha}_1 = \left(\frac{2\sin(\kappa h)}{h}\pm \sqrt{2}A\sqrt{\cos(\kappa h)}\right),
\end{equation}
where $\tilde{\alpha}_1$ is the dispersionless/long-wave speed,
\begin{equation}
    \tilde{\alpha}_3 = -\bigg[\frac{1}{3}h \operatorname{sin}(\kappa h) \\
    \mp \frac{\sqrt{\operatorname{cos}(\kappa h)}}{2 \sqrt{2} A} \left(\operatorname{cos}(\kappa h)-\frac{A^2h^2}{6}\right) \bigg],
\end{equation}
$\tilde{\alpha}_3$ is the coefficient of the leading-order dispersive term, while
\begin{multline}
    \tilde{\alpha}_5 = \bigg(\frac{h^3}{60}\operatorname{sin}(\kappa h) \pm \frac{Ah^4 \sqrt{\operatorname{cos}(\kappa h)}}{960 \sqrt{2}} \\ 
        \mp \frac{h^2\operatorname{cos}(\kappa h)^\frac{3}{2}}{16 \sqrt{2} A } \mp \frac{(\operatorname{cos}(\kappa h))^\frac{5}{2}}{16 \sqrt{2} A^3}\bigg)
\end{multline}
\begin{figure}
    \centering
    \includegraphics[width=0.75\linewidth]{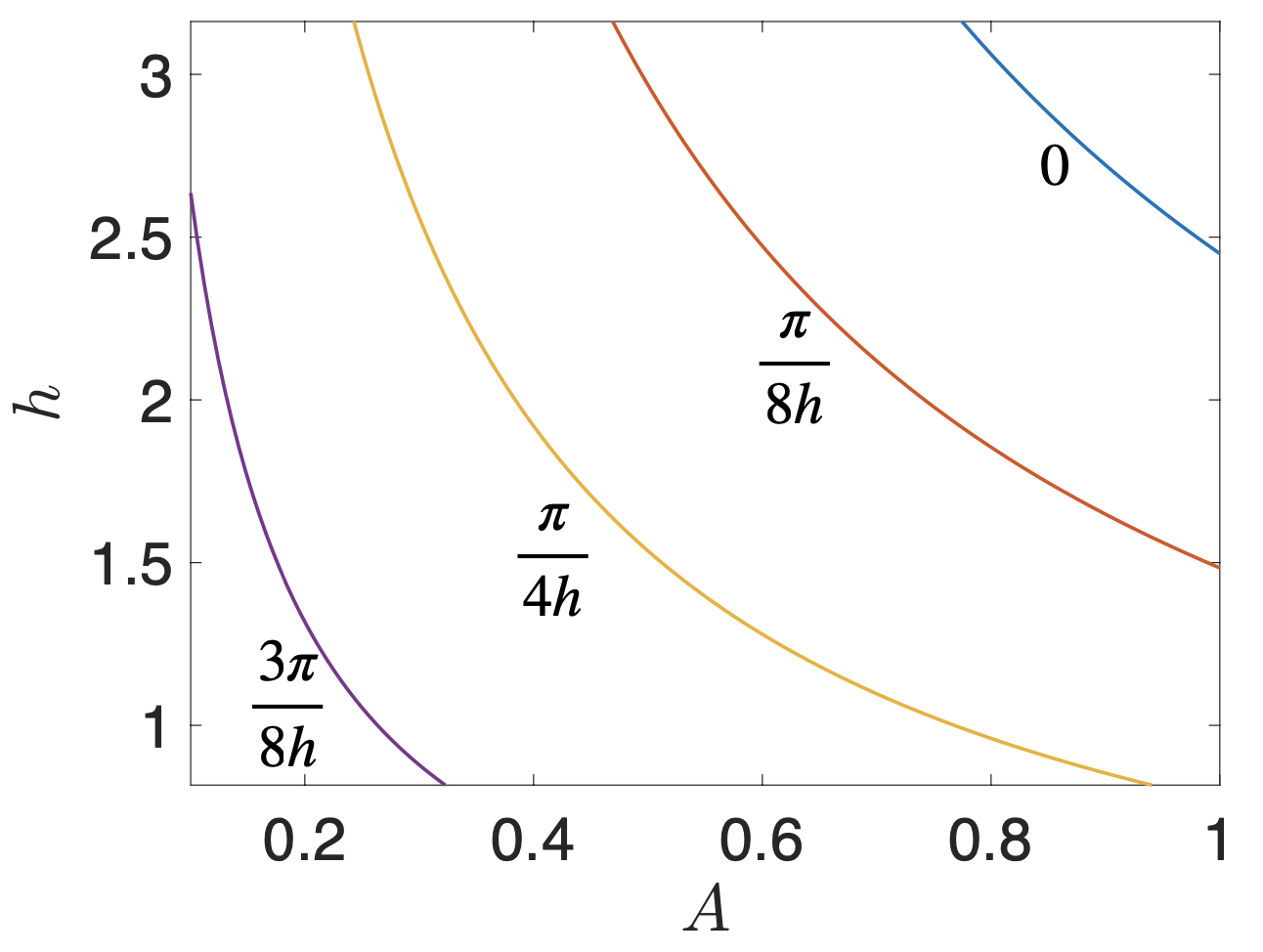}
    \caption{Zero contour map (for various $\kappa$) of the coefficient of the leading-order dispersion ($\tilde{\alpha}_3(A,h;\kappa)$=0). Evidently, for a fixed $\kappa$ and sufficiently large $Ah\gg 1$, a negative leading order dispersion coefficient can be expected.}
    \label{fig:third-order-dispersion}
\end{figure}
represents the coefficient of the $5^{\rm th}$-order dispersion.

Note that the characteristics of the dispersion relation in the long wavelength limit ($K\ll 1$) are dependent on the background amplitude $A$, $\kappa$ (the {``wavenumber" of the Stokes wave}), and the lattice spacing  $h$ (or alternately the coupling strength $\beta$), with the coefficients of the third ({leading}) and fifth-order dispersion terms 
changing signs across different regimes (see, e.g., the contours of 
$\tilde{a}_3$ in Fig.~\ref{fig:third-order-dispersion}), suggestive of (a) competing dispersive effects at different orders and the resulting (b) dispersion curvature sign changes in distinct regimes. 

On  the other hand, for ${A} h \ll 1$, the dispersion relation approaches that of the continuum NLS
\begin{equation}
\label{Continuum_model_dispersion_relation}
    \Omega_{}\pm =2k\kappa\pm 2 k\left(\frac{A^2}{2}+\frac{k^2}{4}\right)^{1/2}.
\end{equation}
In the limit ($h\rightarrow 0$, for any $A$) the dispersion relation in Eq.~\eqref{Disp-relation-band-limited} approaches that of the continuum model (Eq.~\eqref{Continuum_model_dispersion_relation}), with $k\in (-\infty,\infty)$, as was already noted
in~\cite{kivshar1992modulational}.
For the continuum model, the (a) dispersion curvature ($\partial_{kk}\Omega$) is sign-definite and (b) the dispersion is no longer band-limited; both of which represent crucial points of difference between the dynamics of the continuum and discrete defocusing NLS equations.

Note that the dispersion curvature of the discrete NLS is given by
    \small{\begin{align} &\partial_{kk}\Omega_{h,\pm}=-2\sin(kh)\sin(\kappa h)\pm\frac{4}{h}\sqrt{\cos(\kappa h)}h^2\sin^4(hk/2)\times\\\nonumber &\frac{\left(8\beta^2\cos^2(\kappa h)\sin^2(hk/2)\cos(hk)+6\beta A^2\cos(\kappa h)\cos(kh)-A^4\right)}{4\sqrt{2}[\left(\sin^2(hk/2)(2\beta\cos(h\kappa)\sin^2(hk/2)+A^2)\right)]^{3/2}}.
\end{align}}
This quantity can change sign across $k\in [-\pi/h,\pi/h]$.
\section{Simple wave DSW}
\label{Simple-wave-DSW}
In this section we review details necessary to study simple wave DNLS DSW theory through the Whitham-El procedure (see \cite{el_resolution_2005,hoefer2014shock,el2007theory}). We first assume $|u_+|=1>|u_{-}|$, and under the conditions of simple wave theory, the initial step is resolved through a left (right) propagating rarefaction (dispersive shock) wave.

The second (first) Riemann invariant (c.f. Eqs.~\eqref{One-phase-modn}) is constant across the rarefaction wave (DSW). Matching these \textit{jump} conditions produces an expression for the intermediate background density 

\begin{equation}
  u_m=  \sqrt{\rho_m} =\frac{|u_+|+|u_{-}|}{2}
\end{equation}
and an expression for the background velocity (defined implicitly)
\begin{equation}
    E\left(\frac{\kappa_mh}{2},2\right)=\frac{h}{\sqrt{2}}\left({|{u_{-}}|}-|{u_+}|\right).
\end{equation}
$,$ 
We note that the transcendental equation above has no solution past a critical jump of $({|u_{-}|}-|u_{+}|)=\frac{2\sqrt{2}}{h}E(\pi/4,2)\approx \frac{1.2\sqrt{2}}{h}$, representing a definitive limit for the resolution of jumps via rarefaction and DSW's, as predicted by simple wave theory, {for a fixed $h$}. 

Next, below this critical jump, we derive simple wave ordinary differential equations (ODE) from the associated two-phase Whitham modulation equations (c.f. \cite{sprenger2024whitham}) at the linear and solitonic edges that can be used to predict the associated edge speeds of a right-propagating DSW (also referred to as 2-DSW). We first examine this Whitham equation system in the vicinity of the 2-DSW linear edge. 
Here, it is well known that the equation system reduces to a system of three equations given by (see \cite{hoefer2014shock,el_resolution_2005} for a discussion for generalized Schrödinger-type models)
\begin{equation}
    \begin{aligned}
        \left(\overline{\rho}\right)_T + \frac{2}{h}\left(\overline{\rho}\sin\left(\overline{\kappa} h\right)\right)_X &= 0,\\
        \left(\overline{\kappa}\right)_T + \left(\overline{\rho} + \frac{2\left(1-\cos\left(\overline{\kappa} h\right)\right)}{h^{2}}\right)_X &= 0,\\
        k_T + \left(\Omega_{h,\pm}\right)_X &= 0,
    \end{aligned}
\end{equation}
where $\overline{\rho},\overline{\kappa}$ are the averaged hydrodynamic density ($\rho=a^2$) and velocity respectively, over one period of nonlinear dispersive oscillations. It is important
to note here that we consider the evolution over small scales
such that justify the quasi-continuum approach taken above. {Hereafter, we will (without loss of generality) develop the theory for right propagating DSW, i.e. consider only the $\Omega_{h,+}$ dispersion branch.}

We start by attempting to compute the speed of the linear edge. We can obtain $\overline{\kappa}$ through the 2-DSW jump condition \cite{hoefer2014shock} (constancy of the first Riemann invariant)  
    \begin{equation}
        E\left(\frac{\overline{\kappa} h}{2},2\right)=\frac{h}{\sqrt{2}}(\sqrt{\overline{\rho}}-|u_+|).
    \end{equation}
    Substituting this in the evolution equation for $\overline{\rho}$ we obtain, after some simplification,
    \begin{equation}
        \overline{\kappa}_T+\mathcal{V}_h(\overline{\kappa})\overline{\kappa}_X=0,
    \end{equation}
    where $\mathcal{V}_h=\frac{\sqrt{2}}{h}\left(\sqrt{2}\sin(\overline{\kappa} h)+\left(\frac{\sqrt{2}E}{h}+|u_+|\right)h\sqrt{\cos(\overline{\kappa} h)}\right)$.  Notice that in the continuum limit, this reduces to the familiar {dispersionless} velocity $\lim_{h \to 0 } \mathcal{V}_h = \mathcal{V}=3\overline{\kappa} + \sqrt{2}|u_{+}|$.
    
    Our system now reduces to
    \begin{align}
        &\overline{\kappa}_T+\mathcal{V}_h(\overline{\kappa})\overline{\kappa}_X=0,\\\nonumber
        &k_T+(\Omega_{h,+}(\overline{\kappa},k))_X=0
    \end{align}
    We now seek the integral curve $k(\overline{\kappa})$ to this system, which is given by
    \begin{equation}
    \label{Integral-curve-ODE-linear}
        \frac{dk}{d\overline{\kappa}}=\frac{\partial_{\overline{\kappa}} \Omega_{h,+}}{\mathcal{V}_h-\partial_{k}\Omega_{h,+}}.
    \end{equation}
    This is tantamount to the expression given by the well-known
    DSW fitting technique of~\cite{el_resolution_2005}.
Depending on whether the dispersion curvature is positive (or negative), the polarity of the shock wave is such that the linear edge is the leading (trailing) edge. 
Accordingly, the ordinary differential equation needs to be supplemented with an initial condition $k(\kappa_m)=0$ ($k(\kappa_{+})=0$), {which is the boundary condition at the solitonic edge (see \cite{el_resolution_2005} for a discussion).} 
Then it is integrated (numerically) up to the linear edge $k_L=k(\kappa_+)$ ($k_L=k(\kappa_m)$). Subsequently, the linear edge speed for positive (negative) dispersion is calculated using the group velocity formula $\partial_k\Omega_{+}(k_L,\rho_{+}=u_+^2)$ ($\partial_k\Omega_{+}(k_L,\rho_{m})$). 

When the dispersion curvature is negative, (in particular) the difference between the long wave speed and group velocity $\mathcal{V}_h-\partial_{k}\Omega_{h,+}$ (see Eq.~\eqref{Integral-curve-ODE-linear}) could be zero along certain curves $g(k,\overline{\kappa})=0$, thus possibly precluding the existence of integral curves $k(\overline{\kappa})$. Phenomenologically, such resonances between the long wave speed and group velocity are not possible for the continuum NLS, wherein the linear long wave speed is always greater than the {dispersionless speed}.

 To derive the integral curve ordinary differential equation (ODE) at the soliton edge, we define the conjugate angular frequency (c.f. \cite{el_resolution_2005})
    \begin{equation}
        \tilde{\Omega}_{h,+}=-i\Omega_{h,+}(i\tilde{k}),
    \end{equation}
    with the conjugate phase velocity (or soliton phase velocity) given by $s_{}=\tilde{\Omega}_{h,+}/\tilde{k}$.

    Next, we write down the integral curve ODE for the conjugate wavenumber (derived in a similar vein, c.f. \cite{hoefer2014shock,el_resolution_2005} for details) 
    \begin{equation}
    \label{Integral_curve_soliton_edge}
        \frac{d\tilde{k}}{d\overline{\kappa}}=\frac{\partial_{\overline{\kappa}} \tilde{\Omega}_{h,+}}{\mathcal{V}_h-\partial_{\tilde{k}}\tilde{\Omega}_{h,+}},
    \end{equation}
    which needs to be supplemented with initial condition $\tilde{k}(0)=0$ (or $\tilde{k}(\kappa_m)=0$) for DSWs governed by positive (or negative) dispersion curvature. As long as a cavitation point is not attained within the interior of the 2-DSW (where $a^2=\rho=0$), we select the $\tilde{\Omega}_{h,+}$ conjugate dispersion branch for characterization of the 2-DSW soliton edge speed. Having obtained the conjugate wavenumber at the soliton edge, the soliton edge speed is calculated as $s(\tilde{k}(\kappa_m),\rho_m)$ (or $s(\tilde{k}(0),\rho_+)$) for DSW governed by positive (negative) dispersion curvature. Prior to ascertaining the integral curve, we recall sufficient conditions for the 2-DSW admissibility criteria to hold \cite{hoefer2014shock}, which is particularly important to check when the dispersion curvature has a negative sign. These conditions correspond to the monotonicity of the conjugate dispersion relation $\partial_{\kappa}\tilde{\Omega}_{h,+}>0$ as a function of $\overline{\kappa}$
     and the convexity of the conjugate dispersion relation $\partial_{\tilde{k}\tilde{k}}\tilde{\Omega}_{h,+}(\kappa_m,0^{+})>0$. The former monotonicity condition indicates an additional, tighter threshold for the critical jump $|1-u_{+}|$, across which the rarefaction+2-DSW type pattern is observed when the dispersion curvature is negative. At the critical threshold, we have an extremum (for $\tilde{\Omega}_{h,+}$ in $\overline{\kappa}$) at the initial condition to the integral curve ODE in Eq.~\eqref{Integral_curve_soliton_edge}, i.e., $\partial_{\overline{\kappa}} \tilde{\Omega}_{h,+}(\kappa_m(u_{+}),0^{+})=0$.
    
    Similar to the scenario encountered when computing the linear edge speed from Eq.~\eqref{Integral-curve-ODE-linear}, (with the dispersion curvature being negative), the quantity $\mathcal{V}_h-\partial_{\tilde{k}} \tilde{\Omega}_{h,+}$ (see Eq.~\eqref{Integral_curve_soliton_edge}) could be zero along certain curves $\tilde{g}(\tilde{k},\overline{\kappa})=0$, while one tries to calculate the soliton edge speed. This could possibly preclude the existence of integral curves $\tilde{k}(\overline{\kappa})$.

\section{Dynamics of weak DSW}
\begin{figure}
    \centering
    \includegraphics[width=0.9\linewidth]{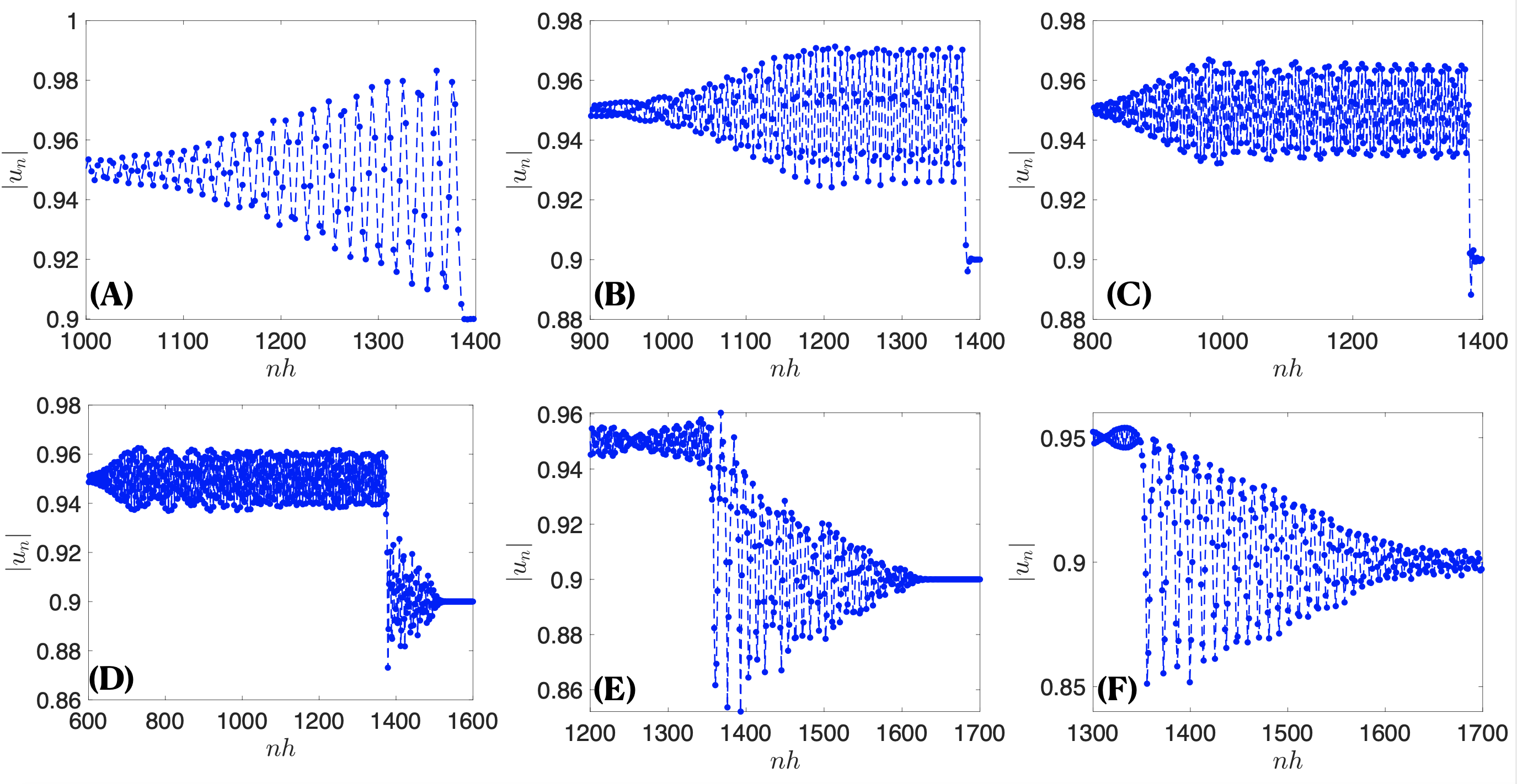}
    \caption{Snapshots of small amplitude DNLS DSW emitted at $t=1000$ for $u_{+}=0.9$, as $\beta$ varies from $0.1$ to $0.6$. Here, in (a) $\beta=0.1$, and we obtain DSW in a regime with negative dispersion curvature $\tilde{\alpha}_3<0$. Thereafter, as $\beta$ is increased, in steps of $0.1$, we observe traveling DSW in (B) and (C) which correspond to $\beta=0.2$, $0.3$ respectively and \textit{crossover} DSW for (D), corresponding to $\beta=0.4$ and finally DSW in a regime with positive dispersion curvature in (E), (F) [$\beta=0.5$, $0.6$ respectively], whose waveforms are significantly influenced by higher-order dispersive effects. }
    \label{fig:catalog_higher_order_dispersion}
\end{figure}

\begin{figure}
    \centering
    \includegraphics[width=\linewidth]{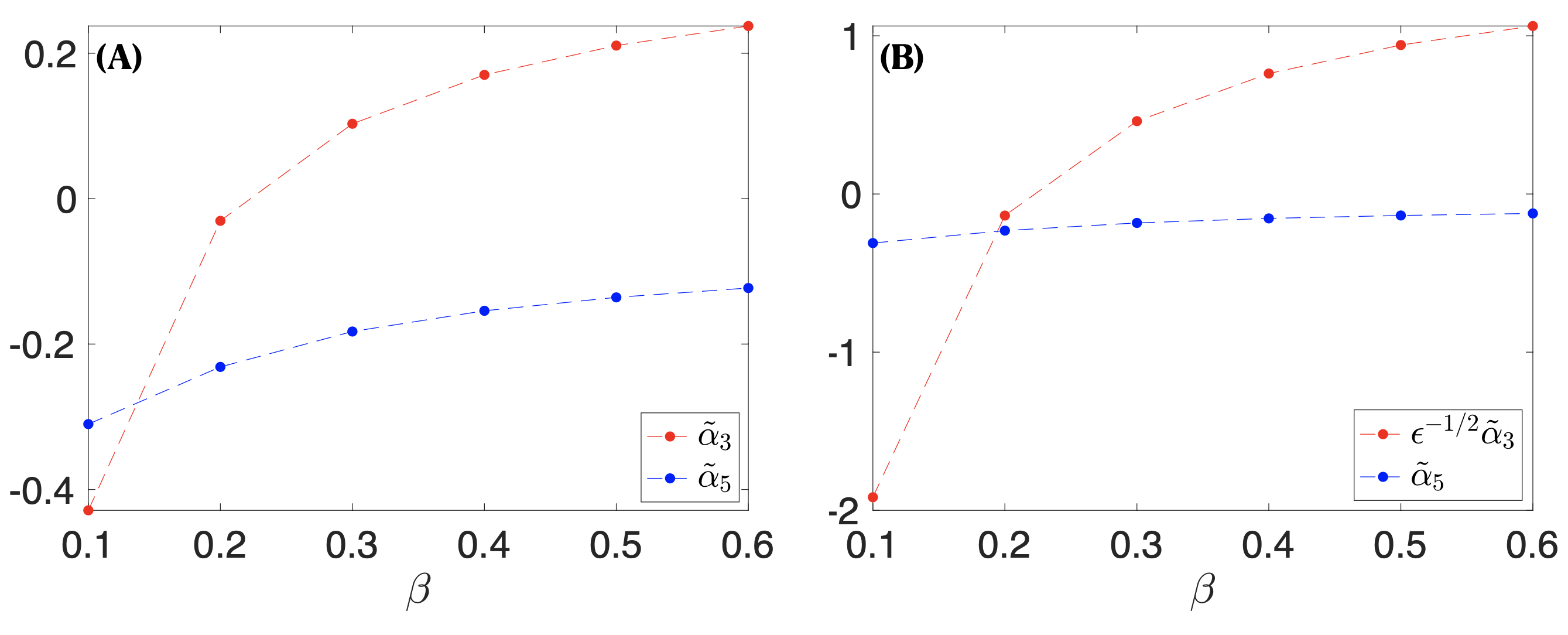}
    \caption{{(A) The dispersion coefficients in the long-wave expansion are evaluated at intermediate hydrodynamic background states ($|u_m|, \kappa_m$), obtained from a sequence of dam-break problems by fixing  $u_{-}=1$, $u_{+}=0.9$ and varying the coupling strength $\beta$. For $\beta=0.1$, the third-order dispersion coefficient is larger in magnitude than the fifth-order term, indicating a regime dominated by negative third-order dispersion. However, near $\beta\approx0.2$, the third-order dispersion coefficient is close to zero signaling the onset of competing dispersive effects at third and fifth orders therein. Eventually for $\beta=0.6$, the third-order term dominates again—this time within a regime of positive dispersion. (B) 
    The (appropriately scaled) third- and fifth-order dispersion coefficients.} }
    \label{fig:alpha3_alpha5}
\end{figure}

\begin{figure}
    \centering
    \includegraphics[width=\linewidth]{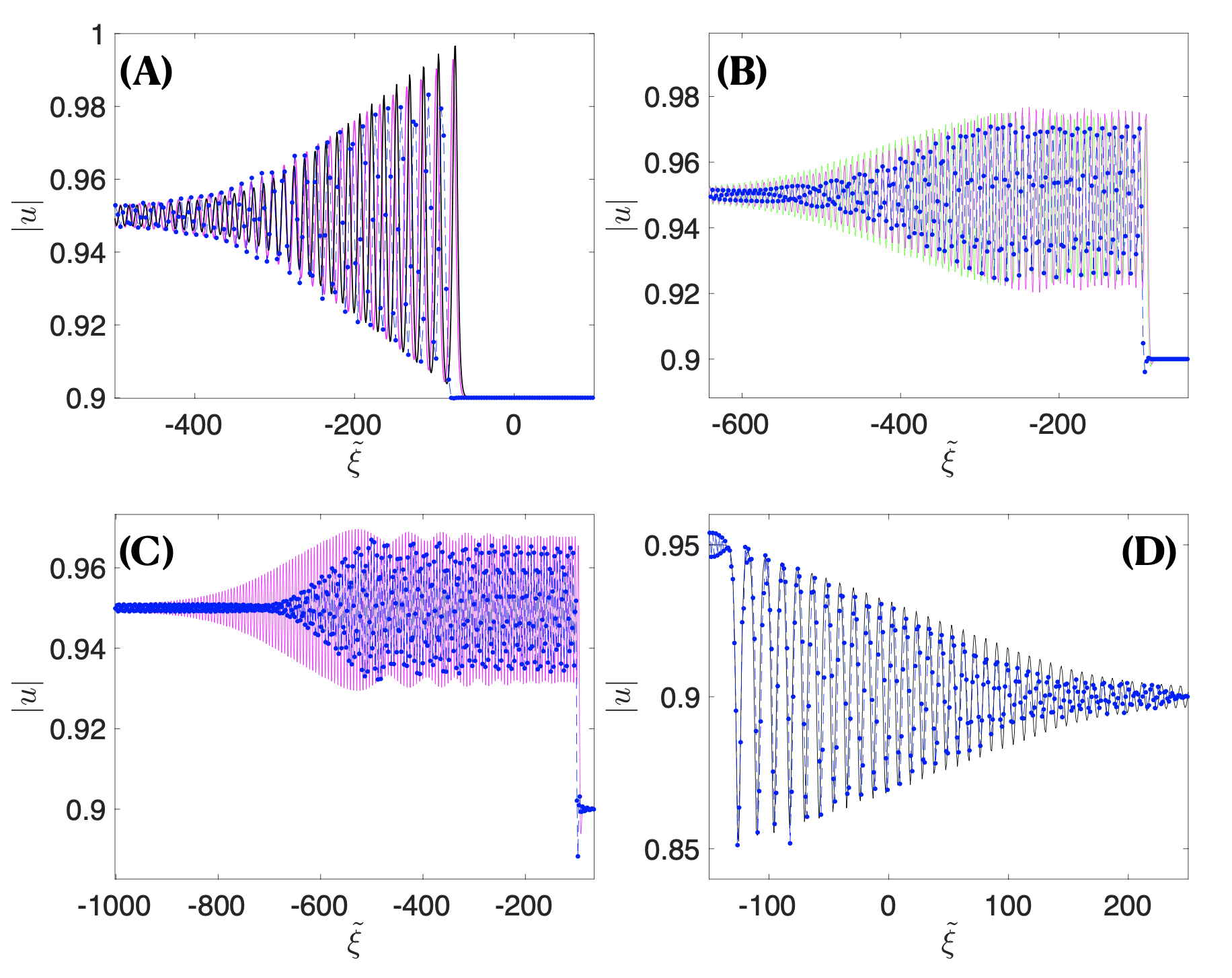}
    \caption{{We compare the right propagating {shock structures} of the DNLS for $\beta=0.1,0.2,0.6$, $u_{+}=0.9$ (snapshot shown for $t=1000$) with corresponding counterparts of the quasi-continuum {Kawahara/KdV5/KdV} reduction depending on their asymptotic validity. The quasi-continuum reductions were derived by examining disturbances propagating on the intermediate background $u_m$ in each case: (A) the DNLS result compared to the Kawahara DSW (magenta) and KdV DSW (black) for $\beta=0.1$, the Kawahara one providing the better approximation, 
    (B) the DNLS result corresponding to $\beta=0.2$ compared to the {KdV5} (green) Kawahara traveling DSW (magenta), with the Kawahara providing the slightly closer approximation, (C) the DNLS result for $\beta=0.3$ compared to its Kawahara reduction (red). Here, the disagreements associated with the linear edge speed (and thus the velocity width) are more pronounced, (D) the DNLS result for $\beta=0.6$ compared to its corresponding KdV reduction. Besides the trailing radiation, good agreement is evident.}}
    \label{fig:Kawahara-DNLS-comparison}
\end{figure}

To investigate the propagation of small (but finite) amplitude disturbances on homogeneous backgrounds, we consider the derivation of quasi-continuum reductions in regimes where \textit{quadratic} nonlinearity balances long-wave dispersion. This is the setting in which the KdV equation and its higher-order weakly nonlinear generalizations typically arise. In this context, quasi-continuum models that go beyond the standard KdV reduction may provide valuable insights into the DNLS dynamics observed for small step sizes, i.e., when $|u_{-} - u_{+}| \ll 1$.

\subsection{Korteweg de-Vries equation}
{
    The Korteweg-de Vries (KdV) is a long-wave, weakly nonlinear model that occurs universally across a variety of scenarios including in fluid mechanics \cite{johnson1980water,Ablowitz2011,whitehead1986korteweg,whitham_linear_1999}, cold atomic gases \cite{kulkarni_hydrodynamics_2012} ion acoustics \cite{roy2008ion,tran1979ion}, besides others. Relevant to the present work, the KdV equation was derived as a quasi-continuum model to describe the observed polarities and dynamics of DSW, besides cataloguing their features within distinct regimes in the well-known Salerno model \cite{konotop1997dark,kamchatnov2004dissipationless}. Here, unit lattice spacing was considered, and variations in the background density parameter was seen to lead to distinct dispersion regimes.
In our work, we observe rich nonlinear-dispersive phenomena also for the DNLS, which represents a special limit of the 
    so-called Salerno model \cite{salerno1992quantum}. Here, variations in the coupling strength $\beta$ were seen to interpolate a range of distinct regimes between positive and negative dispersion curvatures in the long-wave limit. 
    
    The KdV reduction, which arises from a balance of quadratic nonlinearity and third-order dispersion, can be derived by considering the asymptotic balance $a\sim \epsilon$, $\phi_x\sim \epsilon$, $x\sim \epsilon ^{-1/2} \xi$ and $t \sim \epsilon^{-3/2}\tau$. 
    Furthermore, we consider the asymptotic expansions in amplitude and phase
    \begin{align}
        &a\sim \epsilon a^{(1)}+\epsilon^2 a^{(2)}+\cdots,\;\\\nonumber &\phi\sim \epsilon^{1/2}\phi^{(1)}+\epsilon^{3/2}\phi^{(2)}+\cdots.
    \end{align}
    with the travelling coordinate
    \begin{equation}
        \xi=\epsilon^{\frac{1}{2}} (x-\tilde{\alpha}_1 t),  \tau = \epsilon^{\frac{3}{2}} t
    \end{equation}
    Substituting these into the dispersive hydrodynamic equation system in 
    the discrete system for $a$ and $\phi$ obtained 
    from Eq.~(\ref{pert}),
    we obtain the equation relating $a^{(1)}$ and $\phi^{(1)}_{\xi}$ at the leading order(s) 
    \begin{equation}
        \phi_{\xi}^{(1)}=\mp \sqrt{\frac{2}{\cos(\kappa h)}}a^{(1)}
    \end{equation}
    and the KdV equation(s) at the trailing order:
    \begin{equation}
    \label{QCKdV}
        a^{(1)}_{\tau} +\tilde{\beta} a^{(1)} a^{(1)}_{\xi} -\tilde{\alpha}_{3} a^{(1)}_{\xi\xi\xi} = 0
    \end{equation}
    where
    \begin{equation}
        \tilde{\beta} = -Ah\tan(\kappa h)\pm 3\sqrt{2}\sqrt{\cos(\kappa h)},
    \end{equation}
    and $\tilde{\alpha}_3$ is the third-order dispersion coefficient in the long wave expansion in Eq.~\eqref{Long-wave-exp}. The reduction maintains its asymptotic validity as long as $|\kappa|\ll \frac{\pi}{2h}$, $A\gg 0$ and away from the surfaces $\tilde{\alpha}_3(\kappa,A,\beta)=0$ and  $\tilde{\beta}(\kappa,A,h)=0$ respectively.
}
Note the sign of dispersion curvature (coefficient of nonlinearity) of the KdV reductions changes across the surface $\tilde{\alpha}_3=0$ ($\tilde{\beta}=0$). In physical terms, this has implications for the polarity and orientation of the underlying DSW, as was exemplified in \cite{kamchatnov_dissipationless_2004-1,konotop1997dark,konotop2000shock}. To study the quasi-continuum KdV approximations of DSW, we study the Riemann problems posed to Eq.~\eqref{QCKdV} given by
\begin{equation}
    a^{(1)}(\xi,0)=\begin{cases}
        0,;\xi<0\\\nonumber
        -1,\;\xi>0.
    \end{cases}
\end{equation}
Here, the coefficients $\tilde{\alpha}_3$ and $\tilde{\beta}$ are defined at the intermediate background of the DNLS wave-pattern whose amplitude and one-phase wavenumber are defined by $|u_m|$ and $\kappa_m$ respectively.The KdV approximation to the DNLS DSW in the cotraveling reference frame (with speed $\tilde{\alpha}_1$) is then given by $\approx |u_m|+\epsilon a^{(1)}(\epsilon^{-1/2}\xi,\epsilon^{3/2}t_{\rm max})$.

Different, and higher-order weakly nonlinear reductions can be derived also when (a) the coefficient of the third order-dispersion is zero, (b) when the physical scenario involves regimes where the cubic dispersion competes with the trailing order quintic dispersion in the long-wave limit, and (c) when the coefficient of nonlinearity is zero. In (c), third-order dispersion could balance nonlinearity at the next order, leading to the derivation of an mKdV reduction (c.f. \cite{kamchatnov2004dissipationless}), which we shall lay out.
\subsection{KdV-5 reduction}
When $\tilde{\alpha}_3=0$, quadratic nonlinearity could balance fifth order dispersion in the underlying DNLS long wave expansion
for specific choices of the plane wave amplitude $A$, 
leading to the derivation of a KdV-5 reduction (c.f. \cite{kamchatnov_dissipationless_2004-1}). Here, the coefficient of the fifth order dispersion $\tilde
{\alpha}_5$ is evaluated on the surface $\tilde
{\alpha}_3(\kappa,A,\beta)=0$. This surface can be written explicitly in terms of the root for $A^{+}(\kappa,h)=(-2\sqrt{2}\sin(\kappa h)+ \sqrt{6+2\sin^2(\kappa h)})/(h\sqrt{\cos(\kappa h)})$ corresponding to the fast reduction. Now, assuming the asymptotic balance $a\sim \epsilon$, $\phi_x \sim \epsilon$, $x\sim \epsilon^{-1/4}\xi$, $t\sim \epsilon^{-5/4}\tau$, we obtain the expansions:
 \begin{align}
        &a\sim \epsilon a^{(1)}+\epsilon^2 a^{(2)}+\cdots,\;\\\nonumber &\phi\sim \epsilon^{1/2}\phi^{(1)}+\epsilon^{3/2}\phi^{(2)}+\cdots.
    \end{align}
    This leads to the uni-directional (fast) KdV-5 limit given by
 \begin{align}
        a^{(1)}_{\tau}+\tilde{\beta}(A^{+})a^{(1)}a^{(1)}_{\xi}+\tilde{\alpha}_5(A^{+}) a^{(1)}_{\xi\xi\xi\xi\xi}=0.
    \end{align}
The KdV-5 admits a greater family of dynamical shock solutions including traveling DSW, which occur due to a nonlinear resonance of a family of (oscillatory) solitary and periodic waves \cite{hoefer2019modulation}. However,
it is relevant to keep in mind that this reduction
is applicable for specific choices of the plane wave enabling
the non-generic constraint $\tilde
{\alpha}_3(\kappa,A,\beta)=0$.

\subsection{Kawahara reduction}
Across the point of inflection of the linear dispersion relation, dispersive effects at different orders in the long wave limit could compete, leading to the generation of novel waveforms and shock structures. A universal reductive model in this context is the Kawahara equation \cite{kawahara1972oscillatory} which was shown to describe phenomena that arise in a variety of physical contexts, including in capillary-gravity water waves, spin-orbit coupled condensates, and nematic liquid crystals \cite{baqer_nematic_2021,baqer_modulation_2020} besides others (c.f. for instance, \cite{sprenger_shock_2017}). While we observe Kawahara-type dynamics in our lattice simulations quite regularly, to the best of our knowledge, it has not been utilized as a quasi-continuum descriptor of discrete DSW. 

 The model can be derived when  the coefficient of third-order dispersion is small ($\tilde{\alpha}_3\sim \epsilon^{1/2}$). Furthermore, we assume the asymptotic balance $a\sim \epsilon$, $\phi_x \sim \epsilon$, $x\sim \epsilon^{-1/4}\xi$, $t\sim \epsilon^{-5/4}\tau$ and perturbative expansions for the amplitude and phase respectively
        \begin{align}
        &a\sim \epsilon a^{(1)}+\epsilon^2 a^{(2)}+\cdots,\;\\\nonumber &\phi\sim \epsilon^{1/2}\phi^{(1)}+\epsilon^{3/2}\phi^{(2)}+\cdots.
    \end{align}
    At the trailing order, we obtain two uni-directional Kawahara limits given by 
      \begin{align}
        \label{kawahara-reduction}&a^{(1)}_{\tau}+\tilde{\beta} a^{(1)}a^{(1)}_{\xi}-{\epsilon^{-1/2}}\tilde{\alpha}_3 a^{(1)}_{\xi\xi\xi}+\tilde{\alpha}_5 a^{(1)}_{\xi\xi\xi\xi\xi}=0,
    \end{align}
   The Kawahara reduction bridges the two underlying KdV limits corresponding to opposite dispersion curvatures. 
   This enables one to capture the transition in shock-features between the two DSWs with opposing orientations, including traveling DSW and cross-over DSW respectively \cite{sprenger_shock_2017,baqer_modulation_2020,baqer_nematic_2021}. Cross-over DSW in particular are accompanied by prominent generation of linear waves near the ``solitary wave" edge of the DSW. It is however important to be mindful of the limited asymptotic validity of the Kawahara reduction (c.f. \cite{el_radiating_2016}), wherein we must have $\epsilon^{-1/2}\tilde{\alpha}_3\sim\tilde{\alpha}_5$. The Kawahara reduction however does provide major qualitative insight to understanding the change in structure for small amplitude DSW as the coupling strength is varied. 

\section{Classification of wave patterns}
\subsection{wave patterns emitted in the weakly nonlinear limit}
\label{Weak-DSW-pattern}
\begin{figure*}
    \centering
    \includegraphics[width=\textwidth]{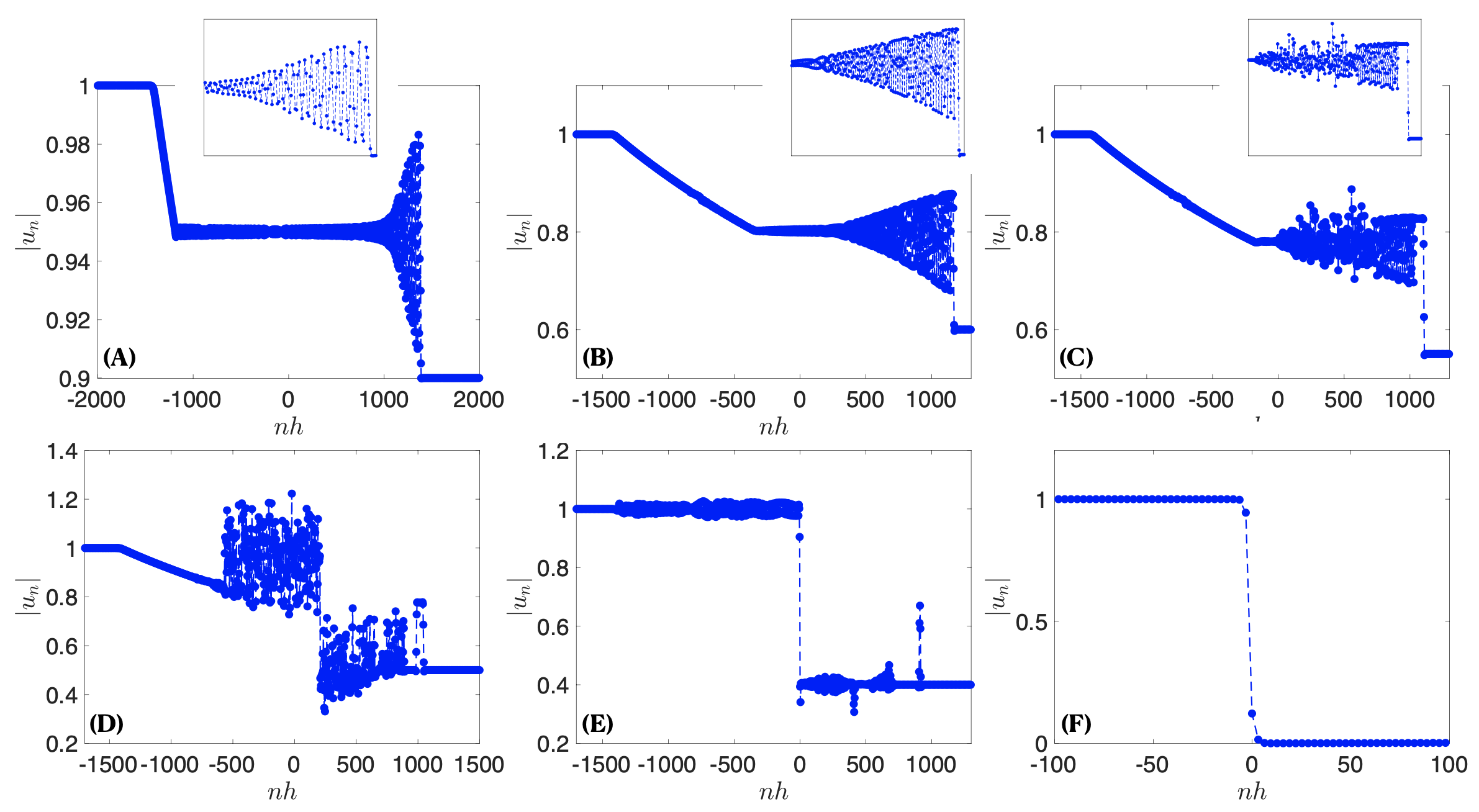}
    \caption{A catalog of DNLS simulations for small coupling $\beta=0.1$ and $u_-=1$ illustrating myriad  wave patterns. The numerical results have been extracted at $t=1000$ (blue dots). We obtain for  (a) small jumps with $u_-=0.9$ being representative; left (right)-propagating rarefaction waves (dispersive shock wave-type features); the polarity of the DSW-type feature is of the bright variety though (associated with negative dispersion curvature), (b) $u_-=0.6$ here a composite right traveling DSW-traveling front are observed to propagate. The DSW appears to resolve an upwards step (in contrast to the cases when $u_->0.6$), (c) the DSW feature breaks down across $u_-=0.6$ ($u_-=0.55$ shown here), (d) For $u_-=0.5$, a train of excitations is seen to be emitted from the DSW breakdown. Notably, box-type, highly coherent {breathers (solitary wave-like)} are emitted, (e) a pulsating feature is formed for $u_-=0.4$ whose breathing leads to the emission of downstream {breathers}, (f) a stationary kink is formed at the location of the initial step when $u_-=0$, which has no analog in the continuum case. }

    \label{fig:catalog}
\end{figure*}
We start our numerical simulations by examining Riemann initial data for which $|u_--u_+|\ll 1$. Key insights are drawn from the weakly nonlinear, {long-wave ($k\ll 1$)} quasi-continuum reductions of the DNLS. To fix ideas, we set $u_-=1$, $u_+=0.9$ for the rest of this section and  examine a collection of Riemann problems with varying $\beta$. We represent the right propagating waves in such a family of Riemann problems in Fig.~\ref{fig:catalog_higher_order_dispersion}. In the weak-coupling limit ($\beta=0.1 \ll1$), we obtain DSWs possessing a polarity corresponding to \textit{bright} solitonic leading edges; see Fig.~\ref{fig:catalog_higher_order_dispersion} (A). For increasing $\beta$, we observe excitations reminiscent of traveling DSW (\ref{fig:catalog_higher_order_dispersion}(B) and \ref{fig:catalog_higher_order_dispersion}(C), where $\beta=0.2,0.3$ respectively). {Traveling DSW are composite waveforms comprising of a \textit{partial} DSW structure and a heteroclinic transition (traveling front) between a periodic state and equilibria. Thus, while they possess the distinct ``linear" and ``solitonic" (which is now non-monotonic)  edges of a DSW, their composite form has a significantly different morphology \cite{sprenger_shock_2017}. } For larger values of $\beta$, e.g., Fig.~\ref{fig:catalog_higher_order_dispersion} (D), $\beta=0.5$) we obtain the analogs of the so-called ``crossover DSW" to the Kawahara equation which represent therein, transitional patterns in a parameter characterizing the relative strengths of the coefficients of leading ($\tilde{\alpha}_3$) -trailing ($\tilde{\alpha_5}$) order dispersion coefficients. {These “crossover” waveforms in the Kawahara context emerge in regimes where non-monotonic solitary waves do not exist. They are distinguished by the presence of both backward- and forward-propagating wave components across the sharp transition toward the upstream equilibrium.} For even larger values of $\beta$, (Figs.~\ref{fig:catalog_higher_order_dispersion} (E)-(F)), corresponding to $\beta=0.5$ and $0.6$ respectively, we essentially observe DSWs with a polarity such that the dark solitonic edge is trailing. There is some trailing-order  in both the aforementioned cases; which in the framework of a Kawahara reduction can be viewed as the \textit{resonance} of the solitonic edge with the spectrum of linear modes. {For further insight on the aforementioned behavior, we show the parametric variations of these long-wave dispersion coefficients in Fig.~\ref{fig:alpha3_alpha5}. Here, it is evident that at the extremes of $\beta$ shown, two different dispersion regimes (negative and positive respectively) are accessed. For these extremal values, a KdV reduction would be expected to provide a reasonable approximation. On the other hand, in the vicinity of $\beta=0.2$, {The coefficient of third-order dispersion in the long-wave expansion approaches zero \textit{uniformly} for all wavenumbers, suggesting that a KdV5 (Kawahara) reduction may be a more appropriate model. In this regime, the nonlinear long-wave dynamics is governed by a balance between quadratic nonlinearity and fifth-order dispersion (KdV5), or by the interplay between quadratic nonlinearity with third- and fifth-order dispersion effects (Kawahara).}

To validate the asymptotic ``correctness" of a Kawahara reduction, the scaled third-order dispersion coefficient is shown against the fifth. Here the size of the step the DNLS DSW resolves ($|u_m|-u_+$) is taken to be the relevant scale $\epsilon$. Evidently, these coefficients are of competing magnitudes only within a small interval in $\beta$, with $\epsilon^{-1/2}\tilde{\alpha}_3$ being sufficiently larger for $\beta\geq 0.3$. It is within this narrow interval of $\beta$ that we expect the quasi-continuum Kawahara reduction to provide a reasonably accurate quantitative description of DNLS shocks.} 

The map of Riemann problems in the limit of small $|u_--u_+|\ll1$ as $\beta$ varies, reveals the phenomenology of competing dispersive effects at successive orders in the long wave expansion of the linear dispersion relation of the DNLS Eq.~\eqref{Long-wave-exp}. Thus, as a first attempt to characterize the phenomenology quantitatively, we invoke the universal Kawahara reduction to the DNLS (Eq.~\eqref{kawahara-reduction}). Here, our small parameter $\epsilon\propto(|u_{m}|-u_+)$, where $u_m$ is the amplitude of the intermediate hydrodynamic background. Furthermore, the coefficients $\tilde{\alpha}_{3,5}\equiv \tilde{\alpha}_{3,5}(\kappa_m,|u_m|)$ are evaluated at the intermediate background amplitude and one-phase wavenumber ($\kappa_m$). Then we pose Riemann problems to this quasi-continuum reduction given by $a^{(1)}(x,0)=0$, for $x<0$ and $a^{(1)}(x,0)=-\alpha\epsilon$ when $x>0$. While $\alpha\sim \mathcal{O}(1)$, we must satisfy $\alpha\epsilon=|u_m|-u_+$. It is important to limit ourselves to the regime of asymptotic validity ($\epsilon^{-1/2}\tilde{\alpha}_3\sim \tilde{\alpha}_5$) of the Kawahara reduction to obtain good quantitative agreement ($\epsilon^{-1/2}\tilde{\alpha}_3\sim \tilde{\alpha}_5$).

We summarize these studies in Fig.~\ref{fig:Kawahara-DNLS-comparison}, at a specific time snapshot $t=1000$ with $\tilde{\xi}=x-\tilde{\alpha}_1t$ is a reference frame co-propagating with the group velocity. In Fig.~\ref{fig:Kawahara-DNLS-comparison} (A), we overlay the DNLS dynamics for $\beta=0.1$ (blue dots) together with those of the Kawahara dynamics (magenta solid line) and KdV dynamics (red solid line), observing good agreements (with the Kawahara equation predicting the self-similar width better than the KdV reduction).

{For $\beta=0.2$, we note $\tilde{\alpha}_3\approx 0$, while sharing the same sign as $\tilde{\alpha}_5$ (Fig.~\ref{fig:alpha3_alpha5} (A)). Thus, we compare both the  KdV5 and Kawahara dynamics (Fig. ~\ref{fig:Kawahara-DNLS-comparison} (B)) to the DNLS results, where evidently, the Kawahara predicts the linear edge speed of the traveling DSW better as compared to its KdV5 counterpart. For even larger values of $\beta$, third- and fifth-order dispersion begin to \textit{compete}, i.e., ${\rm sgn}(-\tilde{\alpha}_3\tilde{\alpha}_5)=1$. Moreover, the system rapidly departs from the regime where the Kawahara reduction remains asymptotically valid. This breakdown is clearly evident at $\beta=0.3$, where a significant discrepancy is observed in the linear edge speed of the traveling DSWs. }

 {As a consequence of this quick departure from the Kawahara regime, we note that we were unable to capture the ``crossover" to the regime with positive dispersion which can be accessed by increasing the $\beta$-parameter. However, the Kawahara reduction still yields a phenomenological map to identify the different types of shock waves one could observe in the weakly nonlinear limit: namely, DSWs of either polarity, traveling DSWs, and crossover DSWs. 
Finally, in Fig.~\ref{fig:Kawahara-DNLS-comparison}(D), for $\beta=0.6$, the dominant third-order dispersion coefficient indicates a regime of positive dispersion. In this case, we observe good agreement between the quasi-continuum KdV reduction and the DNLS model, successfully capturing large-scale features such as edge speeds and the soliton edge amplitude. However, a trailing dispersive tail remains present in the DNLS simulation.}

Obtaining better quantitative quasi-continuum approximations to capture the crossover precisely constitutes a particularly interesting
and fruitful direction for futher research.
This can be done, e.g., either by looking at Whitham-type full dispersion models \cite{binswanger_whitham_2021} or utilizing Pad\'{e} approximants \cite{RosenauFPUTFrenkelKontorova, RosenauDynamicsDenseLattice, RosenauRegularizedNavierStokes, NonlinearKleinGordonPadeApproximation}.

\subsection{Wave patterns in the weak coupling limit}.
\begin{figure}
    \centering
    \includegraphics[width=\linewidth]{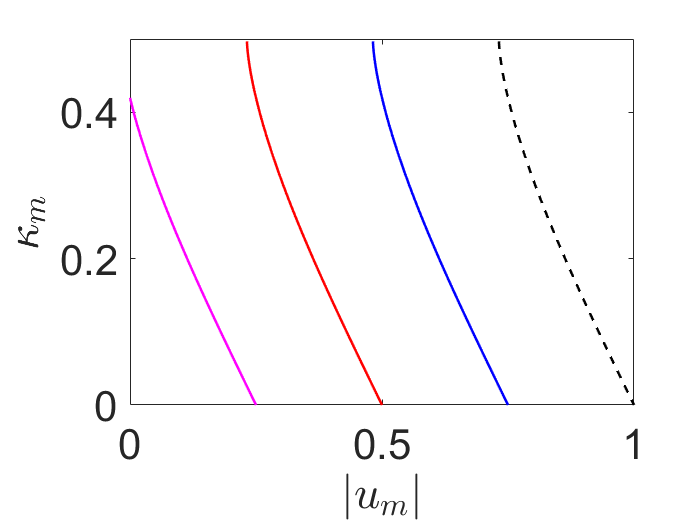}
    \caption{Contours of the second Riemann invariant of the one-phase Whitham modulation equations for $\beta=0.1$. The contours from left (magenta solid) to right (black dashed) denote the levels $0.25$, $0.5$, $0.75$ and $1$ respectively.  The black-dashed isoline represents the possible right-asymptotic states ($|u_m|,\kappa_m$) of the left-propagating rarefaction waves emanating from the Riemann problems whose left asymptotic state is $u_-=1$.}
    \label{fig:rarefact}
\end{figure}

\begin{figure}
    \centering
    \includegraphics[width=\linewidth]{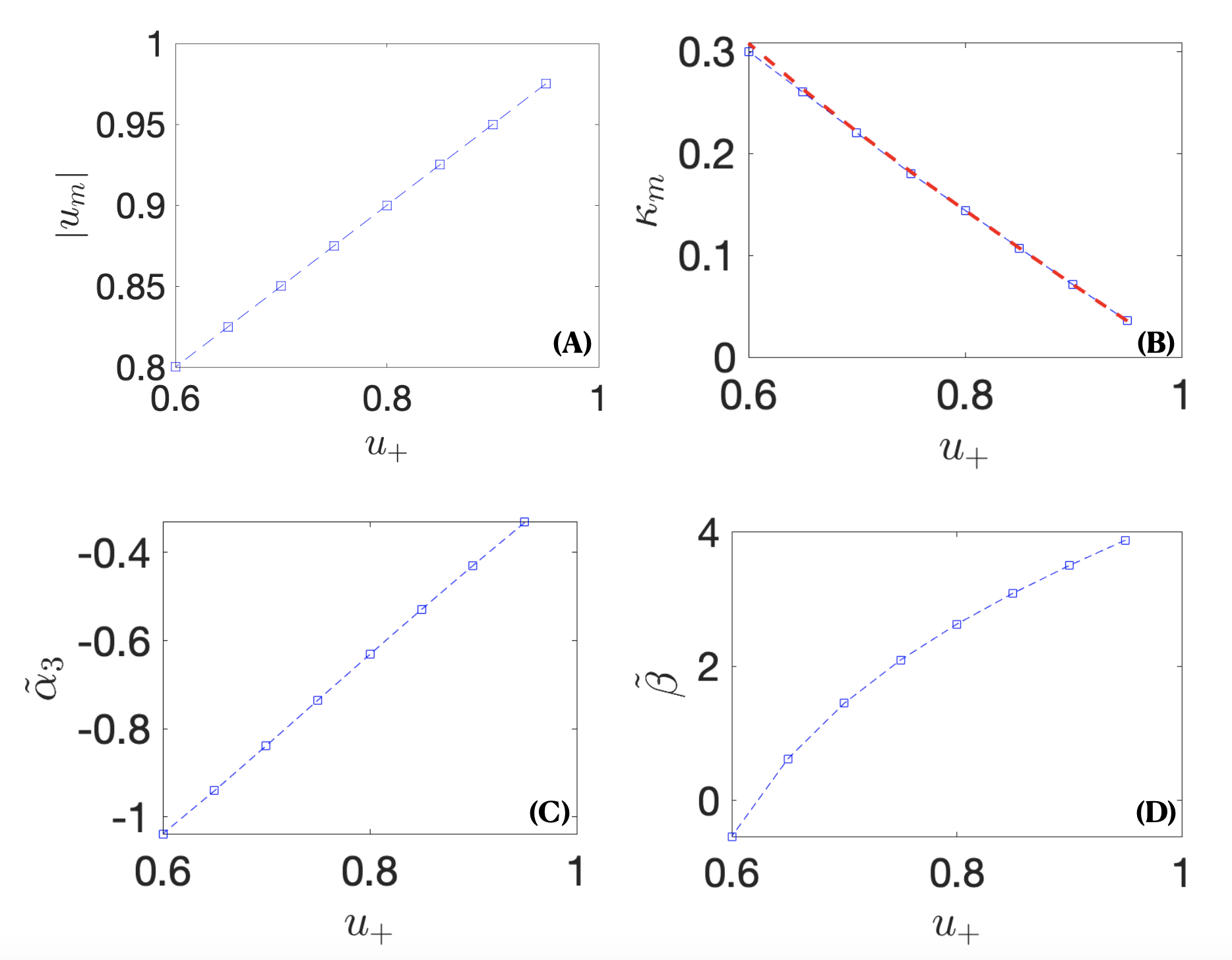}
    \caption{The properties of the DNLS dam breaks in the weak coupling limit $\beta=0.1$ ($u_{-}=1$) as a function of the right hydrodynamic background amplitude $u_+$ for small to intermediate jumps $u_+\in [0.6,1)$. All quantities were extracted at $t=1000$. (A) The numerically extracted (blue squares) for the intermediate hydrodynamic background amplitude agrees  identically with the analytical prediction, (B) The numerically extracted one-phase wavenumber agrees well with the analytically extracted values from simple wave DSW theory (red dashed line) with a slight disagreement around $\kappa_m\approx 0.6$, (C) The numerically obtained third-order dispersion coefficient ($\tilde{\alpha}_3(\kappa_3,|u_m|)$) and (D) the numerically computed nonlinear coefficient $\tilde{\beta}(\kappa_3,|u_m|)$. Note the quantity $\tilde{\beta}$ changes sign close to $u_+\approx 0.62$. }
    \label{fig:PARAMETRIC_PLOT}
\end{figure}

\begin{figure}
    \centering
    \includegraphics[width=\linewidth]{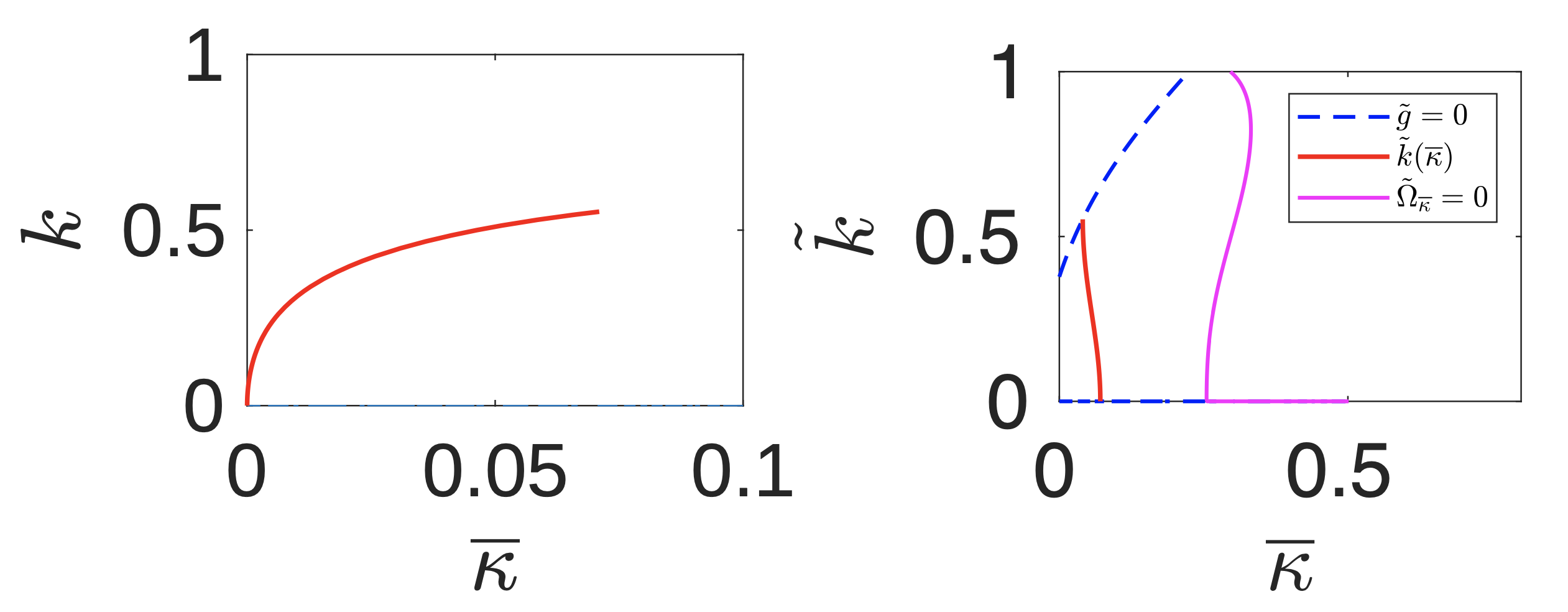}
    \caption{Figures depicting the integral curves (red solid lines) to obtain shock wave edge speeds according to simple wave theory \cite{el_resolution_2005} for the DSW emitted from the Riemann problem characterized by $u_{-}=1$, $u_{+}=0.9$ and $\beta=0.1$. The left panel showcases how the wavenumber is {evaluated} at the linear edge.
    At this value, the group velocity (and thus linear edge speed) is given by $\approx0.9856$. On the other hand, the right panel showcases the failure of simple wave theory at determining the soliton edge speed. Here, the integral curve terminates at the curve $\tilde{g}(\overline{\kappa},\tilde{k})=0$ (see Sec.~\ref{Simple-wave-DSW}). Furthermore, the zero contour of the curve $\partial_{\overline{\kappa}}\tilde{\Omega}$ has been shown in the magenta solid line.}
    \label{fig:Integral_curve}
\end{figure}

\begin{figure}
    \centering
    \includegraphics[width=\linewidth]{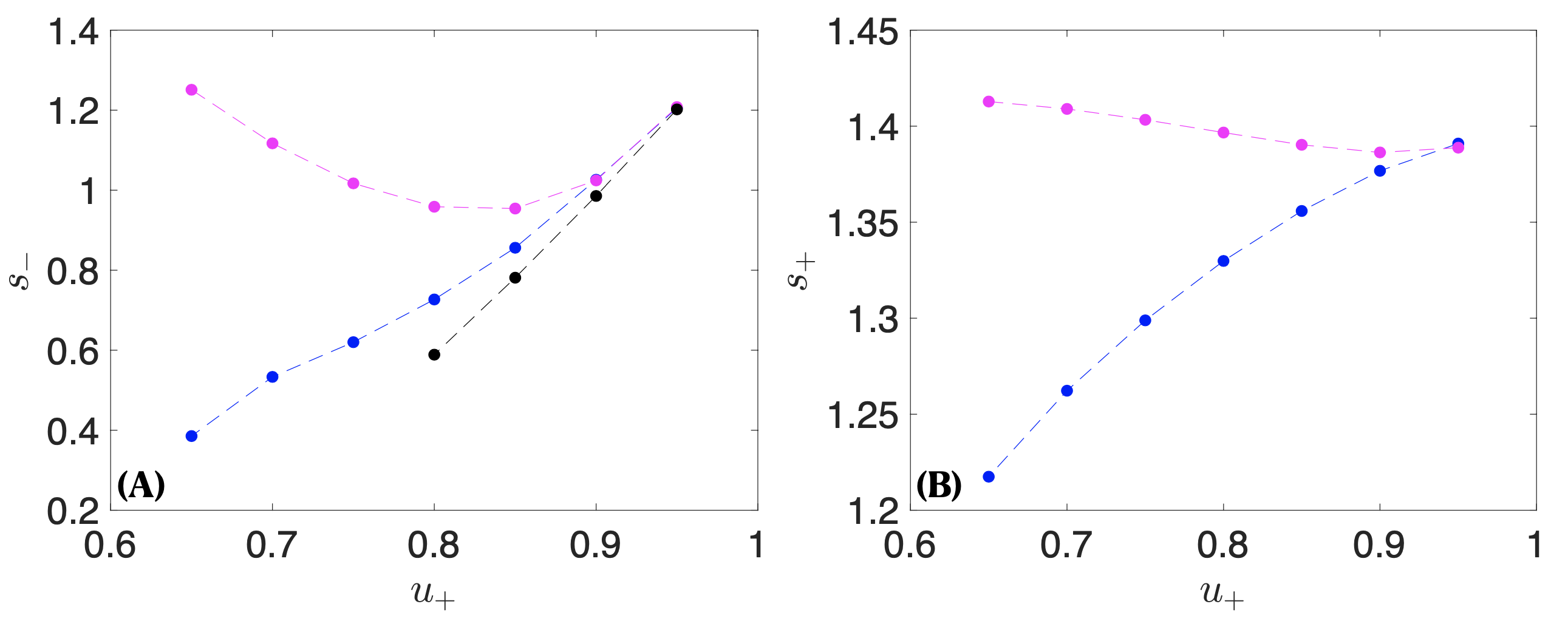}
    \caption{The plots showing the parametric variation of the linear edge ($s_{-}$) and soliton edge ($s_{+}$) speeds as a function of $u_+$ when $\beta=0.1$ and $u_-=1$. The numerical speeds are indicated by the blue dots, the results from DSW fitting (when possible) are indicated by the black dots and the predictions from the KdV reduction are shown using the magenta dots. }
    \label{fig:edge_speeds}
\end{figure}

\begin{figure}
    \centering
    \includegraphics[width=\linewidth]{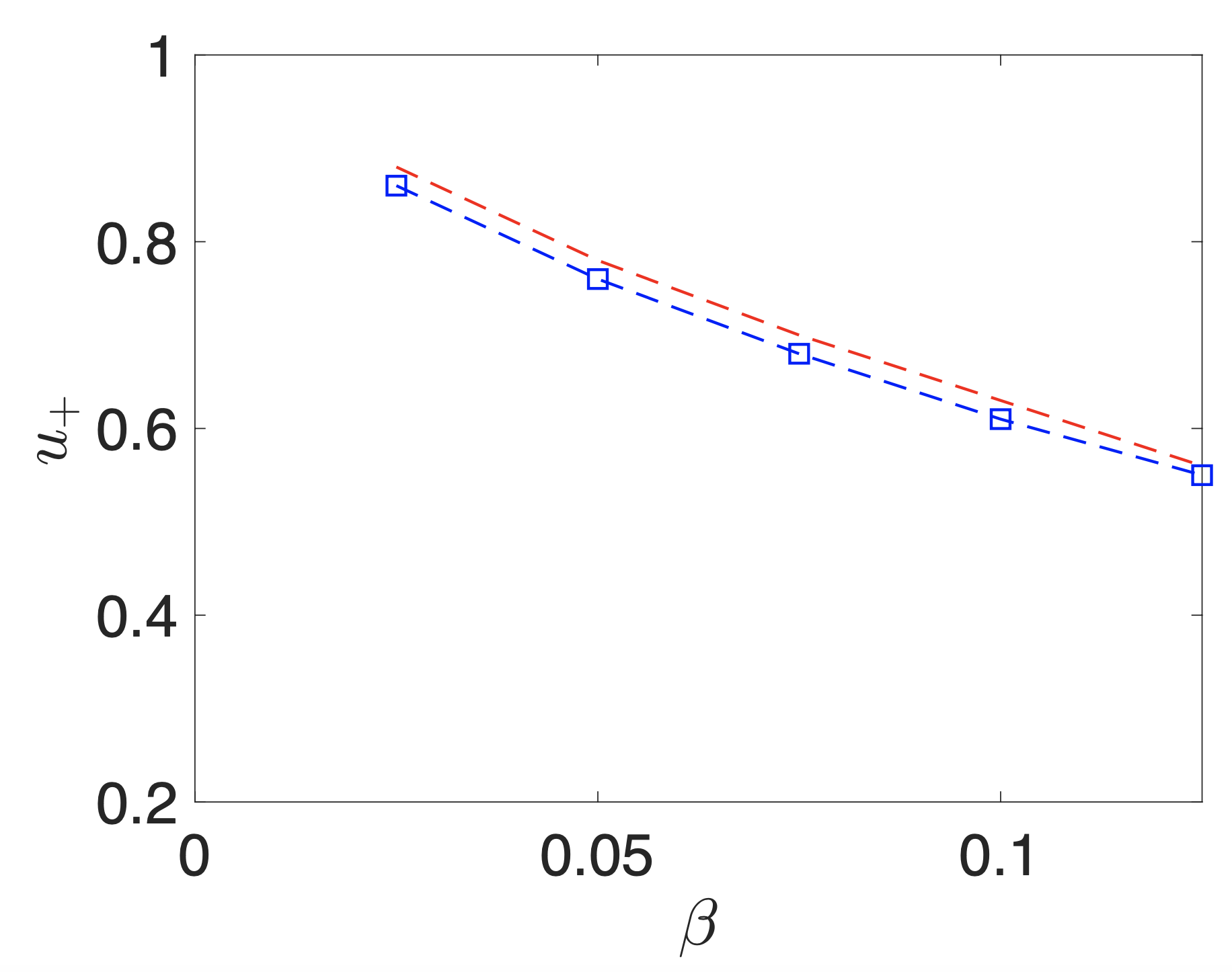}
    \caption{The quasi-analytical (red dashed line) and numerical threshold(s) (blue squares) computed for $u_{-}=1$ and various $\beta$, above which a right-propagating, \textit{composite} wave pattern is expected to be observed.}
    \label{fig:Threshold}
\end{figure}

In this section, we unravel the characteristics of wave patterns that emerge from a family of dam break problems with $u_-=1$, and a representative $\beta=0.1$ while $u_+$ varies in the interval $[-1,1)$. I.e., complementarily to the considerations in the
previous subsection (where the jump height was maintained 
fixed while the coupling $\beta$ was varied), here we
maintain a fixed coupling, while changing the jump height.
In particular, we attempt to study the emitted DSW patterns within the framework of simple wave DSW theory, and we utilize the failure of the theory to corroborate the numerical observations.

To begin our investigations, we first probe the interval of existence (in $u_+$) of the left propagating rarefaction waves for representative small $\beta$. Such rarefaction waves are associated with the first Riemann invariant of the one-phase modulation system in Eq.~\eqref{One-phase-modn}, and thus the second Riemann invariant is constant across their profile. This enforces the
---by now familiar--- transcendental jump condition between their left ($1,0$) and right states ($u_m,\kappa_m$) given by 
\begin{equation}
    (\sqrt{2}/h)E\left(\frac{\kappa_m h}{2},2\right)=1-|u_m|.
\end{equation}
Assuming simple wave DSW theory, $|u_m|=(|u_+|+1)/2$, the equation modifies to $(\sqrt{2}/h)E\left(\frac{\kappa_m h}{2},2\right)=(1-|u_+|)/2$. Clearly, as the right-hand side increases in magnitude (for decreasing $|u_+|$) and given the monotonicity of $E$ (as a function of $\kappa_m$), we cross the hyperbolic-elliptic threshold $\kappa h\geq \pi/2$ eventually. For $\beta=0.1$, this happens when $u_+\approx 0.46$ (or $|u_m|\approx 0.73$: c.f. black dashed line in Fig.~\ref{fig:rarefact}), below which according to the prediction from simple wave DSW theory, rarefaction waves do not exist anymore. This non-existence of rarefaction waves appears to corroborate, at least partially, the dramatic changes that occur within the interval $0.55\leq u_{+}\leq 0.4$. We will discuss this in detail in due course.

At first, we examine the dam breaks when $u_+\in (0.6,1)$. In this regime, the nucleation of a left (right)-propagating rarefaction (dispersive shock wave) across an expanding intermediate hydrodynamic background is observed. Interestingly however, the DSW is seen to possess a polarity of a \textit{bright} DSW since the sign of the dispersion is negative ($\partial_{kk}\Omega_{h,+}<0$) in contrast to the \textit{continuum} NLS in this regime. A representative DSW is shown for $u_+=0.9$ in Fig.~\ref{fig:catalog_higher_order_dispersion}(A), \ref{fig:catalog}(A). We track the variation of the dam break properties in this interval as a function of $u_+$ in Fig.~\ref{fig:PARAMETRIC_PLOT}. We observe perfect agreement between the numerically extracted intermediate wave amplitude $|u_m|$, and the theoretical prediction $|u_m|=(|u_+|+1)/2$ in Fig.~\ref{fig:PARAMETRIC_PLOT}(A); there is a very slight deviation, on the other hand, in the one-phase wavenumber of the intermediate background near $u_+\approx 0.6$ ( Fig.~\ref{fig:PARAMETRIC_PLOT}(B)).  Fig.~\ref{fig:PARAMETRIC_PLOT}(C), showing the sign of the leading order-dispersion coefficient (as a function of the intermediate hydrodynamic background properties) supports the observed polarity of the emitted DSW in this regime. On the other hand, we note a change in sign of the coefficient of nonlinearity ($\tilde{\beta} = -|u_m|h\tan(\kappa_m h)+ 3\sqrt{2}\sqrt{\cos(\kappa_m h)}$) around $u_+\approx 0.62$. This entails a breakdown of the KdV-type reduction (Eq.~\eqref{QCKdV}), {suggests} the non-existence of the bright DSW. The DNLS simulation shows a composite (a DSW and a traveling front) right-propagating wave pattern for values of $u_+\approx 0.6$ (see Fig.~\ref{fig:catalog} (B) ). 

Naturally, the next step in the determination of the macroscopic properties of the dam breaks (in the interval $(0.6,1)$) is the calculation of the linear and solitonic edge speeds using the shock-wave fitting method \cite{hoefer_shock_2014,el_analytic_2005}. To do so, we seek the integral curves associated with the linear edge $k(\kappa)$ and the solitonic edge $\tilde{k}(\kappa)$ by integrating the ODEs in Eqs.~\eqref{Integral-curve-ODE-linear} and \eqref{Integral_curve_soliton_edge} respectively with appropriate initial conditions (see Sec.~\ref{Simple-wave-DSW}). 

The DSW fitting is shown for a representative case $u_+=0.9$ and $\beta=0.1$. The associated linear (solitonic) edge integral curve is shown in the left (right) panel of Fig.~\ref{fig:Integral_curve}. $k(\kappa_m)$ in the left panel yields the wavenumber at the linear edge, and the associated edge speed is given by computing the group velocity $\partial_k\Omega_{h,+}(\kappa_m,\rho_m)$, where $\rho_m=|u_m|^2$. The linear edge speed for this representative example agrees with the numerical edge speed up to within $7\%$. However, the attempt to apply the fitting procedure to calculate the solitonic edge speed fails; see the right panel of Fig.~\ref{fig:Integral_curve}, where the integral curve  $\tilde{k}(\kappa)$ is seen to collide with the curve of singularity $\tilde{g}(\overline{\kappa},\tilde{k})=0$ (see Sec.~\ref{Simple-wave-DSW}). Thus, the integral curve fails to exist past the collision, which explains the failure of the fitting procedure to calculate $\tilde{k}(0)$. The DSW fitting procedure (for $\beta=0.1$) does not yield a prediction for the solitonic edge speed over a wide range of $u_+\in(0.6,1)$ due to this resonance between the long wave speed and the conjugate group velocity. Such a resonance between the long wave speed and the group velocity, i.e., existence of zero-contours $g(\overline{\kappa},k)=0$ can also occur in the $k$-$\kappa$ plane; referred to as \textit{two-phase resonance} \cite{sprenger2024whitham}. The DSW fitting procedure in 
{this} case 
{fails to yield the linear edge speed}: this is something we have observed in our simulations for $\beta=0.1$ when $u_{+}<0.8$. 

An alternative procedure is the utilization of the weak DSW theory to give predictions on the edge speeds (see Sec.~\ref{Weak-DSW-pattern}), {which become increasingly accurate as the jump height $1-u_+$ (and consequently $|u_m|-u_+$) is reduced}. To do so, we compute the edge speeds associated with the bright DSW of the KdV reduction in Eq.~\eqref{QCKdV}. This is known to yield analytical formulas for the edge speeds given by $s_{-}= 2\tilde{\beta}(u_+-|u_m|)+\tilde{\alpha}_1$ (linear edge speed) and $s_{+}=\frac{1}{3}\tilde{\beta}(u_+-|u_m|)+\tilde{\alpha}_1$ (solitonic edge speed) \cite{el_analytic_2005}.

We summarize the results of such comparisons between the numerically computed edge speeds and those through DSW fitting (when possible) and also the KdV reduction edge speeds in Fig.~\ref{fig:edge_speeds}. The comparisons indicate that the regime of \textit{weak} DSW in this extreme regime ($\beta\ll 1$) is when $u_+\lessapprox 0.9$. Furthermore, we see that the DSW fitting at the linear edge appears to capture the decreasing trend in edge speed (see Fig.~\ref{fig:edge_speeds} (A)), while is is applicable. On the other hand, the KdV reduction linear edge speed fails to capture even the \textit{trend} in edge speeds for $u_+<0.9$: this appears to be indicative of beyond KdV-effects in this regime associated with small coupling.

As we mentioned previously, around $u_+\lessapprox 0.6$, we no longer observe the right propagating \textit{bright} DSW pattern (cf. Fig.~\ref{fig:catalog} (B)): instead, a composite DSW-traveling front pattern is observed. Furthermore, the orientation of the DSW in this composite pattern is such that it appears to resolve a \textit{step-up} Riemann problem. To obtain further analytical insight into the change in behavior across this threshold in $u_+$, we turn to the DSW fitting method, wherein we have a sufficient condition for 2-DSW admissibility given in Sec.~\ref{Simple-wave-DSW}, which also suggests a threshold for their existence. Thus, at the critical threshold for the existence of 2-DSW, we must have $\partial_{\overline{\kappa}}\tilde{\Omega}_{h,+}(\kappa_m(u_+),0^{+})=0$. We have shown this threshold for a representative range of $\beta$ in Fig.~\ref{fig:Threshold} (red dashed line) and compared this to the numerically determined threshold (by direct simulations), demonstrating good agreement.
\begin{figure}
    \centering
    \includegraphics[width=\linewidth]{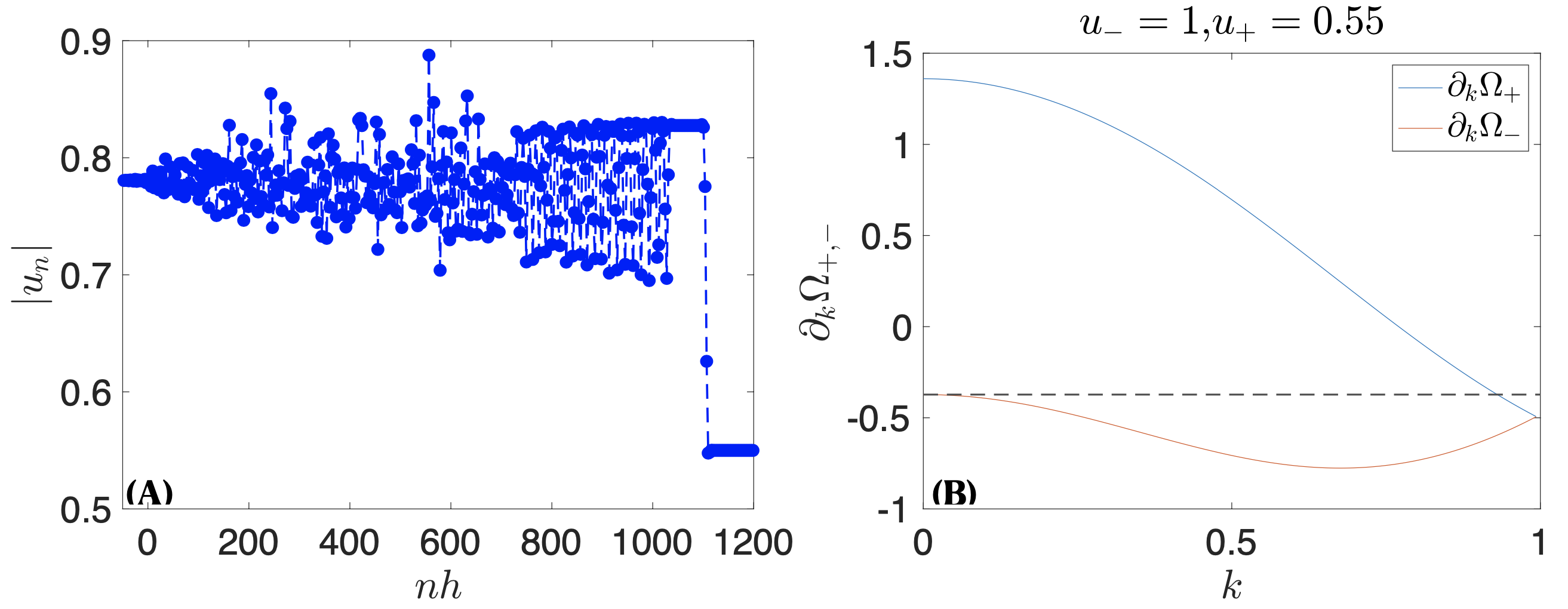}
    \caption{ The breakdown of the DSW in the composite waveform is observed  just across $u_{+}<0.6$, when $\beta=0.1$ and $u_{-}=1$. (A) A snapshot of the dynamics at $t=1000$ for the representative Riemann problem with $u_-=0.55$, depicting the breakdown, (B) Resonance between the dispersionless limit of the $\partial_{k}\Omega_-(|u_m|,\kappa_m)$-branch and a short wave associated with the $\partial_{k}\Omega_+(|u_m|,\kappa_m)$ branch observed at the DSW linear edge in this case.}
    \label{fig:two-phase-resonance}
\end{figure}

Yet another important threshold corresponds to the observation of DSW breakdown in the composite wave pattern for $\beta=0.1$ and $u_+\leq0.59$ (c.f. Figs.~\ref{fig:catalog} (C), \ref{fig:two-phase-resonance} (A) where the dynamics associated with $u_+=0.55$ is shown).{Curiously, just beyond this threshold, a resonance between the dispersion branches emerges at the linear edge $(|u_m|,\kappa_m)$ of the DSW feature. Specifically, the dispersionless speed $\mathcal{V}(|u_m|,\kappa_m)\equiv\partial_k\Omega_{-}(|u_m|,\kappa_m)|_{k=0}$ coincides with the short-wave speed of the second dispersion branch, i.e., $\partial_k\Omega_{+}(|u_m|,\kappa_m)|_{k\approx \pi/h}$. Within the framework of weakly nonlinear modulations of two-phase waves, and under mild assumptions of a constant leading-order wave mean $(|u_m|,\kappa_m)$, it has been shown that such a two-phase resonance constitutes a mechanism leading to modulational instability (MI) of small-amplitude periodic (Stokes) waves \cite{sprenger2024whitham,whitham_linear_1999}. This mechanism for MI is qualitatively distinct in that it predicts a \textit{singularity} to the induced wave mean terms at trailing order (and in the nonlinear frequency correction of Stokes waves). Importantly, this can occur even in the regime of modulational stability of one-phase waves/homogeneous backgrounds.} {To sense the possible presence of such an instability in our simulations, we monitored the Fourier spectrum over the time interval preceding the onset of instability in the DNLS dam breaks (for a range of $u_+\leq 0.59$) and observed the excitation of short-wavelength modes resonant with the long-wave group velocity associated with the $\partial_k \Omega_-(0; |u_m|, \kappa_m)$ branch. However, to definitively attribute the observed instability (Fig.~\ref{fig:catalog}(C), \ref{fig:two-phase-resonance}(A)) to the phenomenon of two-phase resonance, it is necessary to revisit the analysis of small-amplitude Stokes wave modulations in \cite{sprenger2024whitham} and extend that towards the study of the short-time development of the instability.} Finally, we mention that the phenomenon of two-phase resonance is impossible for the \textit{continuum} defocusing NLS.

{The nonlinear stage of instability is witnessed to lead to the generation of a highly oscillatory region characterized by multiple \textit{phases}, something also mentioned in an earlier work \cite{panayotaros2016shelf}, and also, the generation of a train of excitations downstream to the breakdown (c.f. Fig.~\ref{fig:catalog} (D)). 
A systematic characterization of the relevant, apparently
higher-genus phenomenology is presently unavailable, to the best
of our knowledge.}

\begin{figure}
    \centering
    \includegraphics[width=\linewidth]{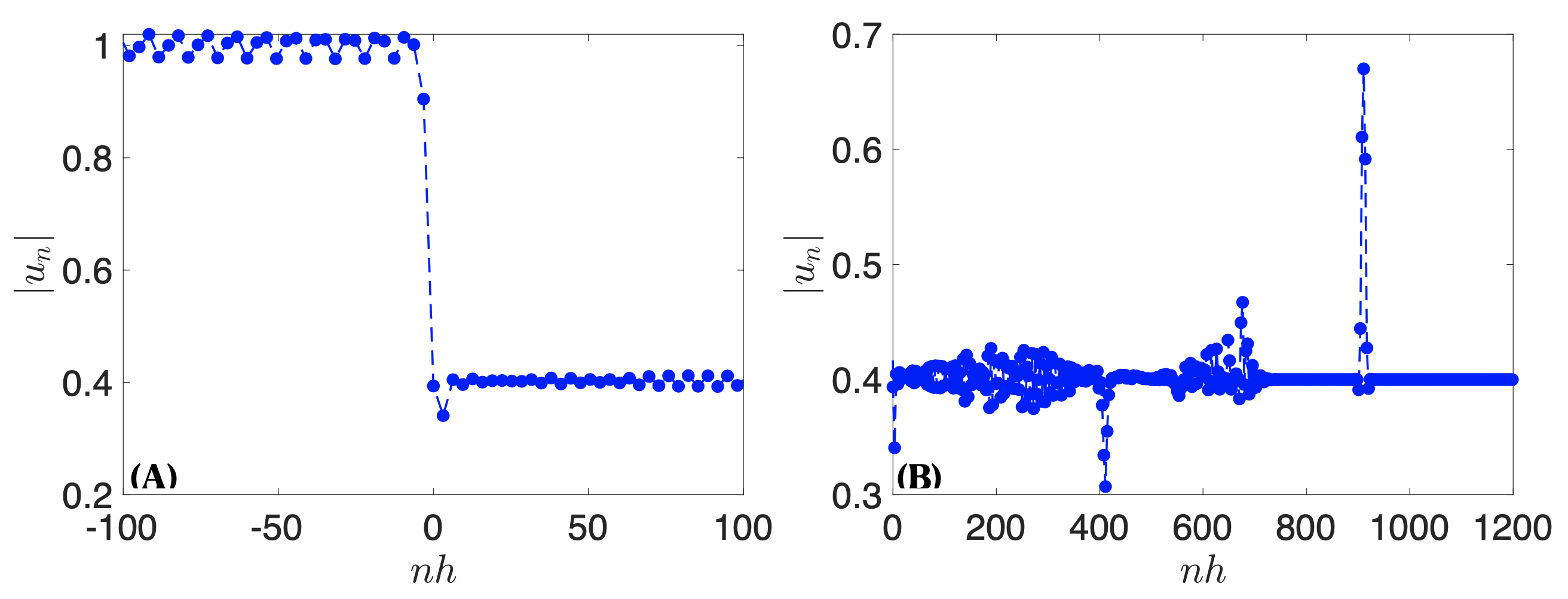}
    \caption{The wave pattern resulting from a Riemann problem with $u_+=0.4$ for $\beta=0.1$ (A) Unsteady heteroclinic transition (central sites) between a periodic wave and equilibria, (B) Train of small amplitude right propagating excitations.}
    \label{fig:Heteroclinic}
\end{figure}

\begin{figure}
    \centering
    \includegraphics[width=\linewidth]{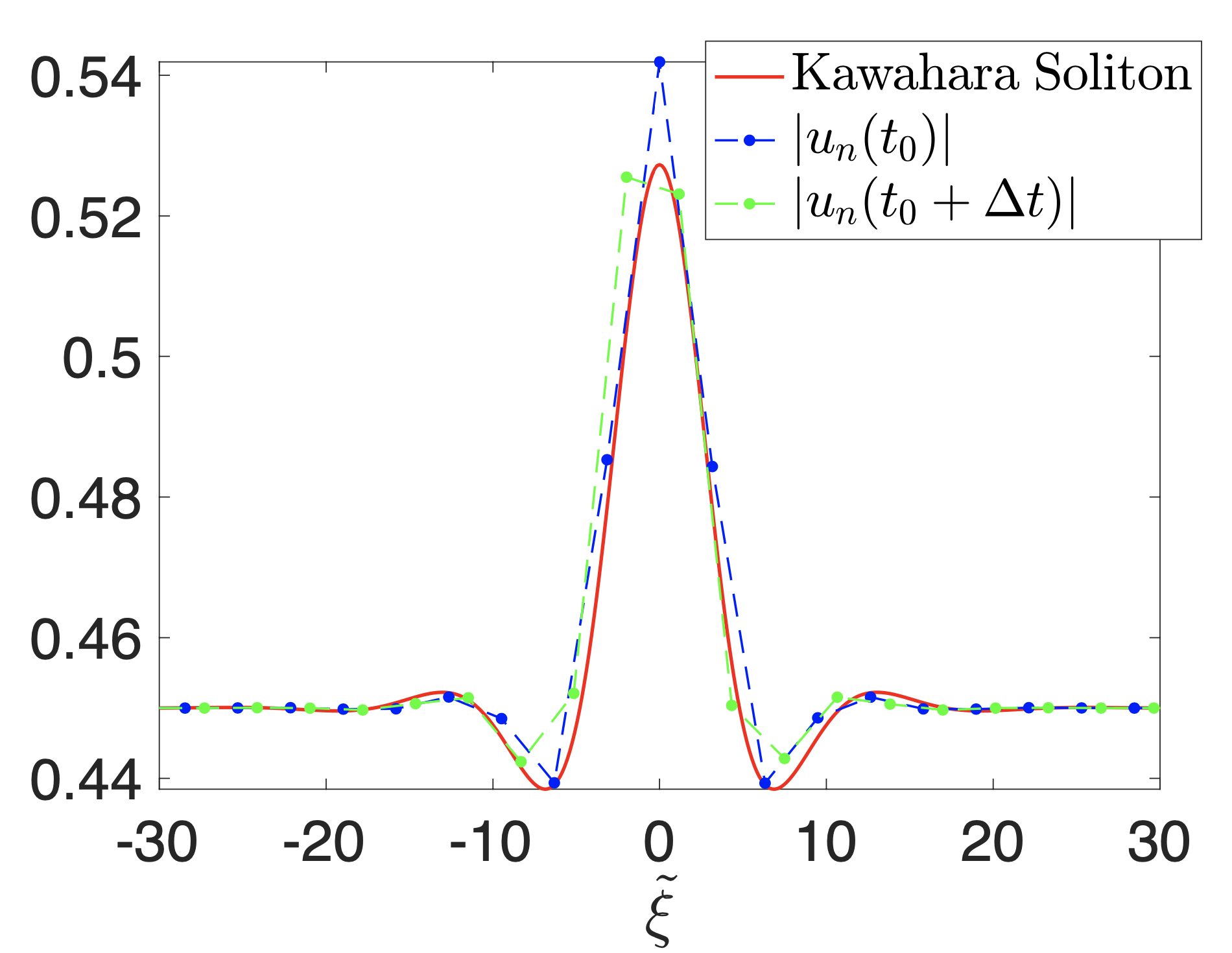}
    \caption{Comparison of two different realizations of a small amplitude DNLS ($c\approx 0.79$) breather in the co-traveling frame with an approximating Kawahara soliton. The traveling breather was emitted from a Riemann problem with $u_{+}=0.45$, $u_{-1}=1$, with both of its most distinct realizations still ``closely" approximated (maximum relative error in $L_{\infty}$ norm across snapshots is to within 4$\%$) by the co-moving Kawahara soliton.}
    \label{fig:Kawahara_soliton_DNLS_breather_comparison}
\end{figure}

 For sufficiently small values of $u_+$ 
 (c.f. Fig.~\ref{fig:catalog} (E), $u_+=0.4$), the multi-phase structure generated due to generalized modulational instability transitions to an unsteady \textit{heteroclinic} transition (across the central sites) between a two-phase wave and a nearly homogeneous state on the right (Fig.~\ref{fig:Heteroclinic} (A)). The train of right propagating excitations carries far less ``mass'' (squared $l^2$ norm) and has two prominent waves. The first is a wave that dips in intensity below the background, and is embedded within the radiation; the second prominent excitation is a \textit{bright} feature propagating faster than the radiation (c.f. Fig.~\ref{fig:Heteroclinic} (B)). The localized bright feature, in particular, appears to pulsate slightly during its propagation, a feature that cannot be discerned in the
 still of Fig.~\ref{fig:catalog} (E) and Fig.~\ref{fig:Heteroclinic} (B). 

To obtain some analytical insight into the bright excitation (observed frequently for $u_+<0.5$), we isolate a representative \textit{ small amplitude} ($|a|\ll u_+$) mode emitted from a Riemann problem characterized by $u_+=0.45$ and $\beta=0.1$. The pulsating small amplitude excitation is reminiscent of oscillatory Kawahara solitons \cite{kawahara_oscillatory_1972,sprenger_shock_2017}. We perform such a fit of a numerically computed Kawahara soliton \cite{yang2009newton} in the co-traveling frame $\tilde{\xi}=x-ct$ (Fig.~\ref{fig:Kawahara_soliton_DNLS_breather_comparison}). Here, two singificantly different realizations of the traveling breather are compared to the Kawahara soliton (c.f. Eq.~\eqref{kawahara-reduction} whose coefficients defined by the background $u_+$). We find that the Kawahara soliton approximates the traveling ``breather" reasonably well with the relative error across snapshots limited to $4\%$. Several such small-amplitude traveling ``bright" breathers appear to be well approximated by the Kawahara solitons. The larger-amplitude (and faster) variants  of such bright excitations have a distinct box-type profile (cf. Fig.~\ref{fig:catalog} (D)).  To reveal some of the properties
of this box-state, we follow its center of mass using the formula {$x_{\rm COM}$ $=$ $\bigg(\sum_{k = 1}^{N} \frac{1}{\sqrt{\beta}} k |u_k|^2\bigg)$ $\bigg(\sum_{k = 1}^{N} |u_k|^2\bigg)^{-1}$}. A linear fit to the center of mass function unveils its approximate speed (c.f. Fig.~\ref{fig:DNLS_center_of_mass_in_box_type_breathers} (A)) to be $\approx 1.047$. The periodic pulsations of this breather in the co-moving frame can be probed by computing the Fast Fourier transform of the periodic time series $x_{\rm COM}-v_{\rm COM} t$. This reveals the dominant frequency $\approx 1/3$ (Fig.~\ref{fig:DNLS_center_of_mass_in_box_type_breathers} (C)). This calculation is verified directly by time evolution of the waveform up to $t=3$, which reveals the waveform reverts to its original shape (Fig.~\ref{fig:DNLS_center_of_mass_in_box_type_breathers} (D)). The complete characterization of this family of breathers, most notably
in the context of the observed box waves is yet another intriguing
feature worthy of consideration for further work.
Indeed, this is technically a rather intriguing topic as 
it involves the potential for traveling states within the DNLS model,
a feature that has been long debated in numerous 
earlier works~\cite{GOMEZGARDENES2004213,Baras}.

 \begin{figure}
     \centering
     \includegraphics[width=\linewidth]{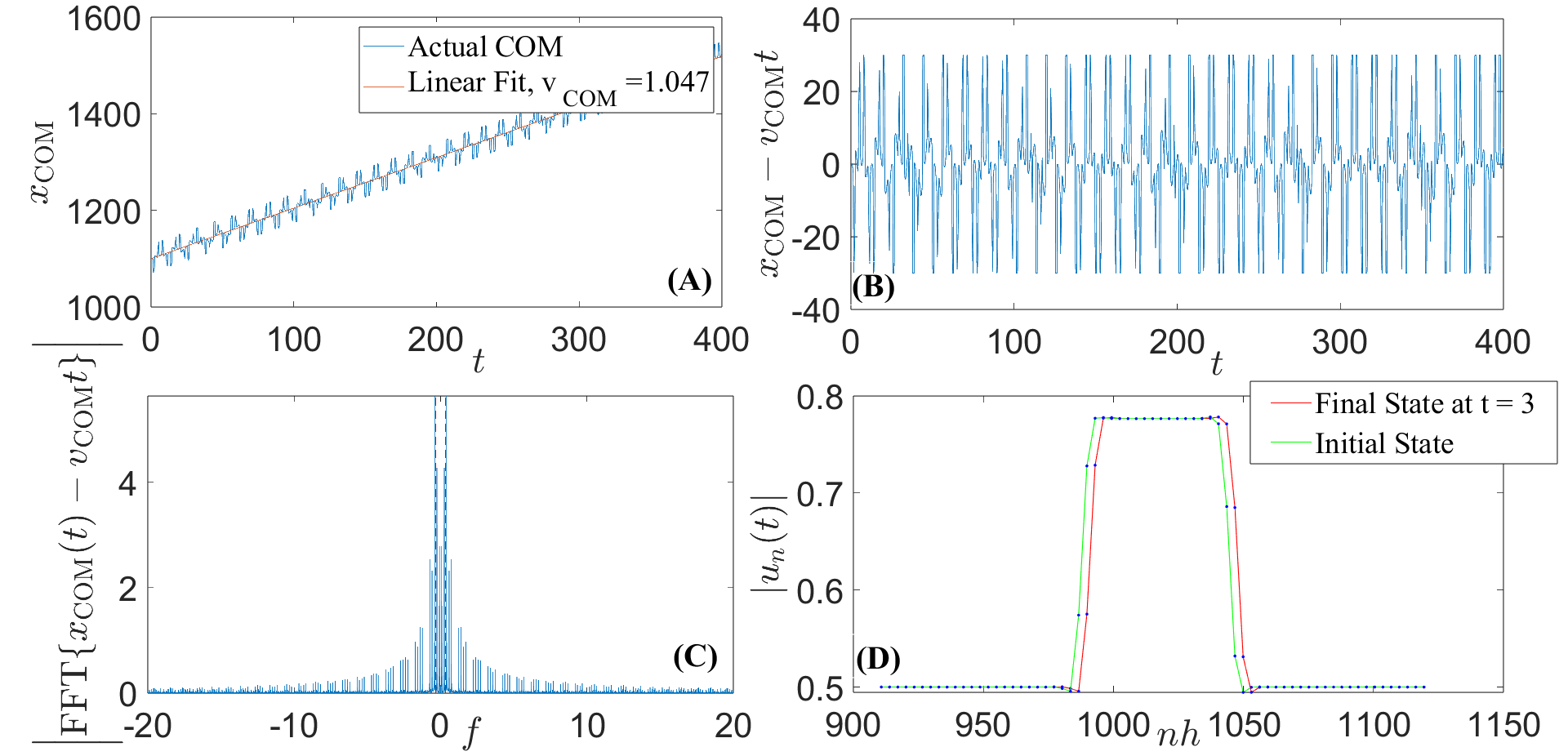}
     \caption{ (A) The center of mass together with its linear fit reveals the velocity of the box-type breather is $v_{\rm COM} = 1.047$. 
     (B) The breathing pulsations of the box-breather as a function of time, (C) Spectral content of the oscillations with the dominant frequency around $1/3$. (D) Here, a comparison of the waveforms at $t = 0$ and $t = 3$ is suggestive of the approximate time period of the breather pulsations in the co-traveling frame.}
     \label{fig:DNLS_center_of_mass_in_box_type_breathers}
 \end{figure}

For $u_+=0$ ($\beta=0.1$), the unsteady heteroclinic in Fig.~\ref{fig:Heteroclinic} (A) merges into a stationary \textit{kink} feature, which has no analog in the continuum NLS case. This novel feature is interesting in its own right, and we discuss its properties in a separate Sec.~\ref{VACUUM-DAM-BREAK} below.
\begin{figure}
    \centering
    \includegraphics[width=\linewidth]{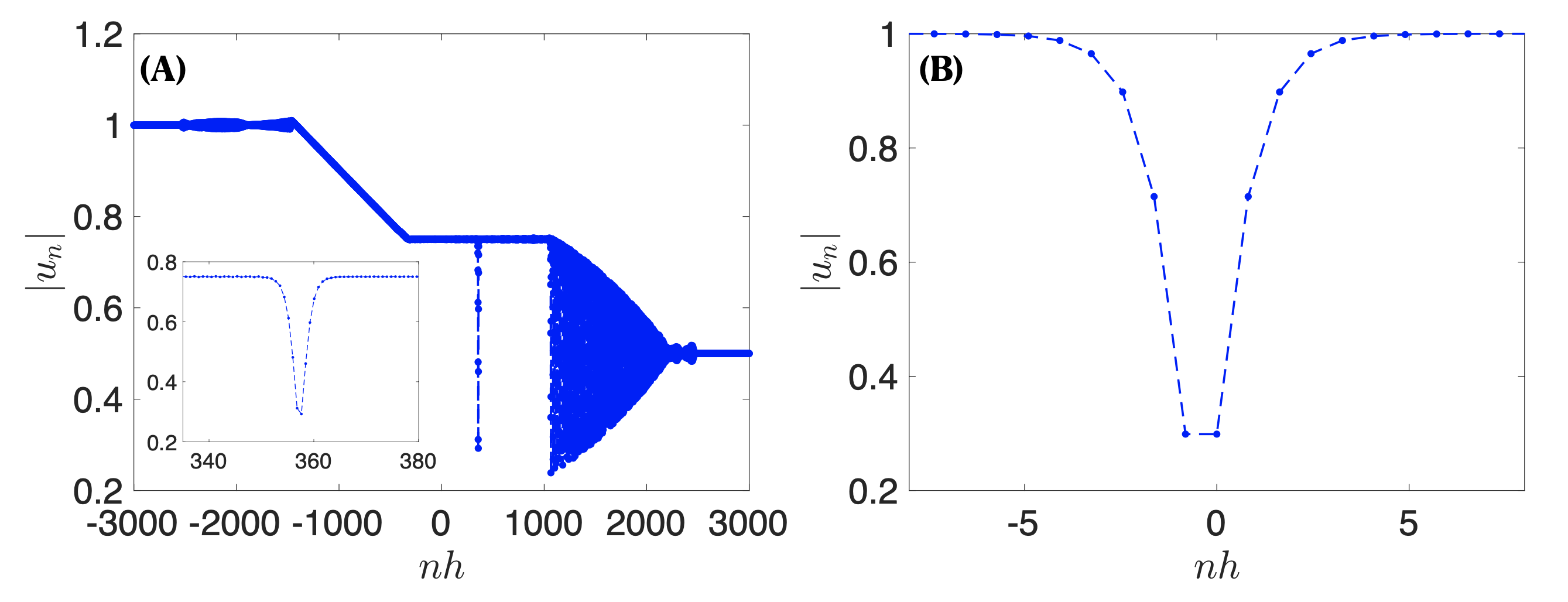}
    \caption{Representative wave patterns that emerge for $u_+<0$ above the large-coupling threshold, i.e. $\beta=1.5>\beta_c$. (A) For $u_+=-0.5$, a left (right) propagating rarefaction (DSW) are observed, whose structure and properties are identical to when $u_+=0.5$. The initial \textit{phase} jump between the equilibria $u_{\pm}$ is mediated by the formation of a gray-solitonic like feature (see also inset). (B) When $u_+=-1$, the phase jump across the equilibria leads to the formation of an intersite dark soliton. }
    \label{fig:continuum_DNLS}
\end{figure}

\begin{figure*}
    \centering
    \includegraphics[width=\linewidth]{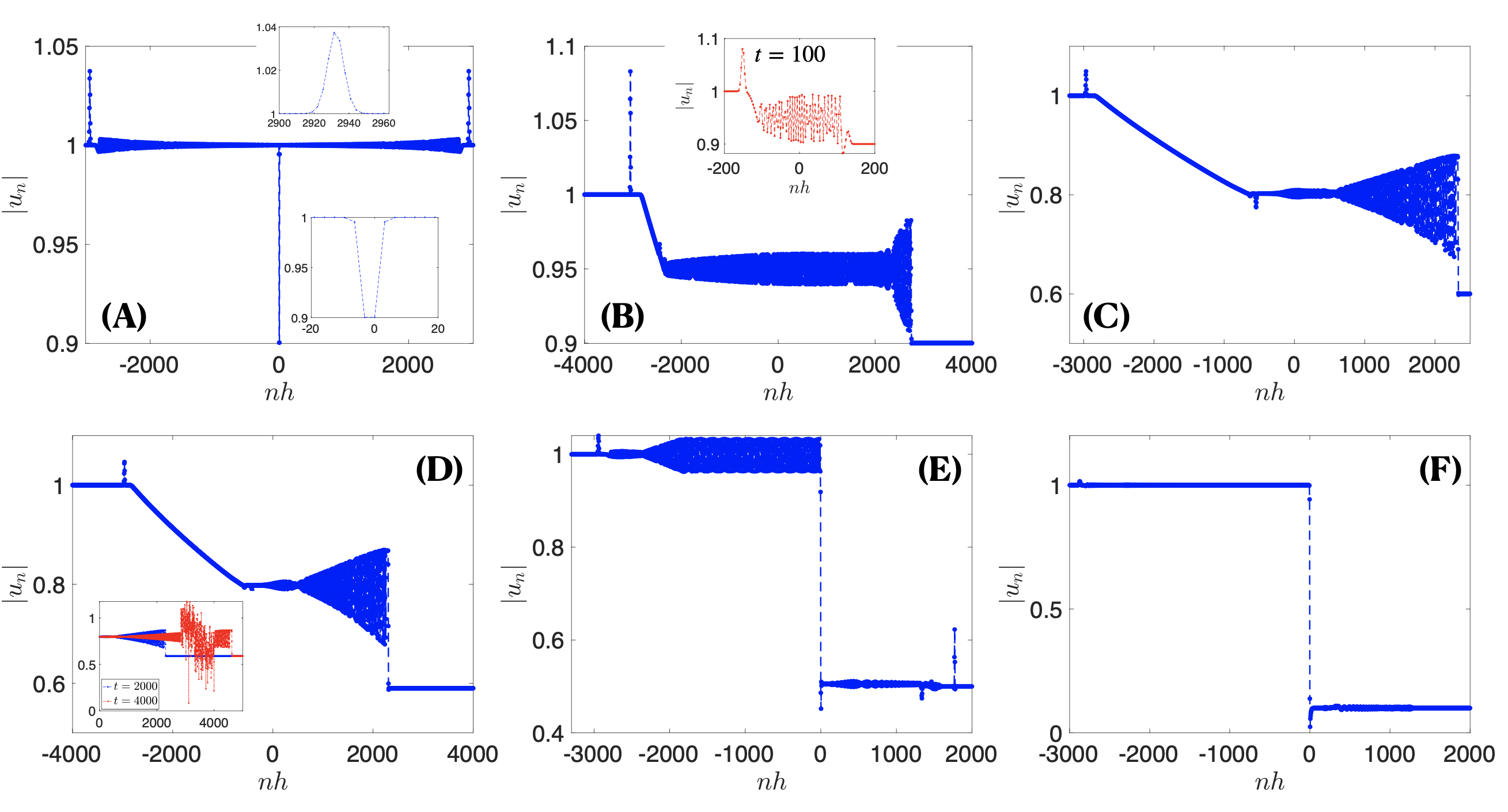}
    \caption{Snapshots of wave patterns at $t=2000$ for $u_+<0$ in the small coupling regime $\beta=0.1\ll \beta_c$. The initial phase jump plays a more dramatic role in differentiating the wave-patterns with $u_+>0$ here, than when $\beta\geq \beta_c$: (A) For $u_+=-1$, a intersite DS is observed in the central sites (akin to the large $\beta$ case), however, counterpropagating moving solitary waves are also generated, (B) For $u_+=-0.9$, the development of the rarefaction-DSW pattern (see inset for $t=100$) is significantly impacted (contrast to Fig.~\ref{fig:catalog}(A)), (C) For $u_+=-0.6$, the right propagating composite wave pattern (also seen when $u_+>0$, Fig.~\ref{fig:catalog}(B)) is observed, (D) For $u_+=-0.59$, the {suggested} two-phase resonance acts toward DSW breakdown in the composite wave pattern over long times (see inset), (E) For $u_+=-0.5$, a heteroclinic connection between a periodic state (left) and equilibria (right) is witnessed. This feature has a jump in phase across the central lattice sites in the same sense as the initial condition (Fig.~\ref{fig:neg_phase_jump_heteroclinic}), (F) The heteroclinic feature is seen to persist over a wide interval in $u_+\in[-0.5,0)$, a representative case corresponding to $u_+=-0.1$ is shown. {The space-time dynamics of this breathing feature is shown in Fig.~\ref{fig:1_0n1}(A) and (B).} }
    \label{fig:neg-values}
\end{figure*}

\begin{figure}
    \centering
    \includegraphics[scale = 0.4]{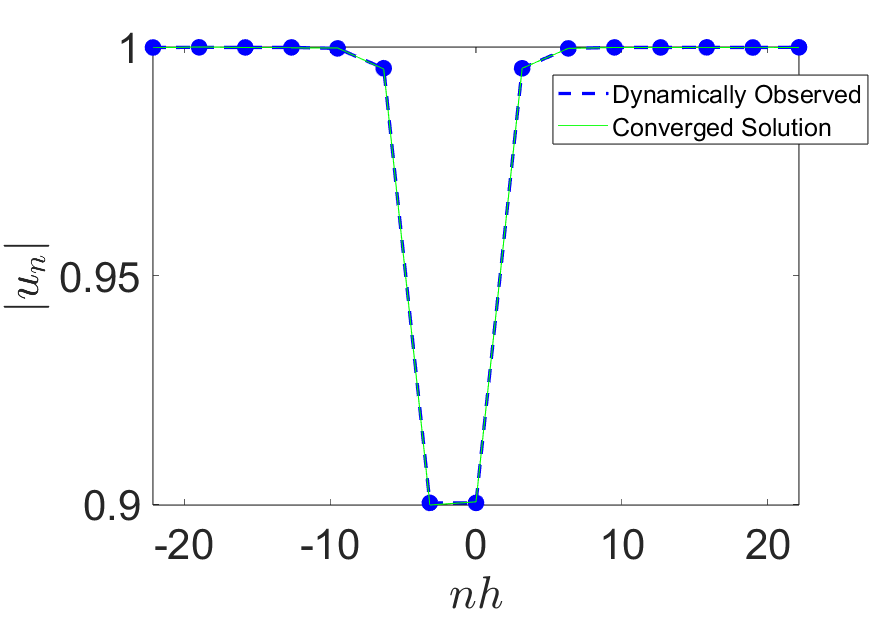}
    \caption{The dark inter-site soliton observed in the Riemann problem dynamics with $u_+=-1$ and $\beta=0.1$. A snapshot of the dynamics at $t=2000$ is compared to a fixed-point solution to the discrete NLS displaying good agreement.}
    \label{fig:DarkSoliton}
\end{figure}

\begin{figure}
    \centering
    \includegraphics[width=\linewidth]{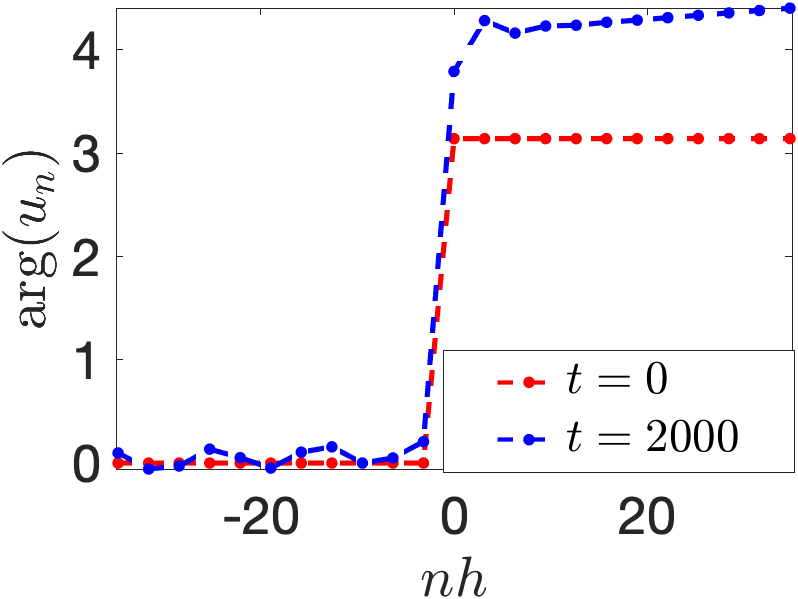}
    \caption{The snapshot of the dynamics (in ${\rm arg}(u_n)$) is shown at $t=0$ (red dots) and $t=2000$ (blue dots) for the Riemann problem with $u_+=-0.5$ and $\beta=0.1$. The polarity of the initial phase jump is preserved by the heteroclinic transition, whose amplitude variation was depicted in Fig.~\ref{fig:neg-values} (E). }
    \label{fig:neg_phase_jump_heteroclinic}
\end{figure}

Next, we examine the cases corresponding to $u_+<0$. The initial $\pi$-phase jump across the central sites has a prominent influence on the dynamics in the weak-coupling limit $\beta<\beta_c$. We contrast the strong- and weak-coupling limits to shed light on this aspect. We display snapshots of the dynamics at $t=2000$ in the strong coupling limit ($\beta=1.5>\beta_c$) [c.f. Fig.~\ref{fig:continuum_DNLS}]. Here, in Fig.~\ref{fig:continuum_DNLS} we display a snapshot of the dynamics at $t=2000$ for $u_+=-0.5$ (A) and for $u_+=-1$ (B) respectively. In (A), the wave pattern in the far field (away from the central sites) is identical to when $u_+=0.5$, with the phase jump between the equilibria mediated by the formation of a gray-solitonic like feature. This waveform radiates energy as it propagates. On the other hand, an intersite dark soliton (DS) is seen to arise across the central sites in (B) ($u_+=-1$), with 
decaying in amplitude, outward propagating, bi-directional radiation. 

We have now set the stage for the examination of the corresponding dam breaks in the weak-coupling limit. Snapshots of the dynamics at $t=2000$ have been shown in Fig.~\ref{fig:neg-values}. For $u_+=-1$ (Fig.~\ref{fig:neg-values} (A)), besides the intersite DS formed at the central sites, counterpropagating ``bright" moving 
(indeed, practically traveling) {breathers} are generated. 
The observed moving structures here are nearly solitary wave-like, and in the given dispersion regime $\tilde{\alpha}_3(|u_+|,0)<0$ may be approximated by KdV (Eq.~\eqref{QCKdV}) solitons of the form $|u_+|+a_s {\rm sech}^2\left(\sqrt{\frac{a_s}{12}}\sqrt{-\frac{\tilde{\beta}}{\tilde{\alpha_3}}}\epsilon^{1/2}(x-(\tilde \alpha_1+V_s\tilde{\beta}\epsilon )t)\right)$. We find that these KdV solitons for a given amplitude are narrower in extent. Perhaps the inclusion of higher-order dispersion could lead to a closer agreement.
We 
have also seen that the dynamically observed DS compares well with a solution obtained by solving a boundary value problem for the discrete NLS utilizing Newton iterations \cite{carretero2025nonlinear}. Excellent initial guesses for the Newton iterations leading to a stationary DS can be generated in the weak-coupling limit by considering the perturbation expansions $v_n = v^{(0)}_n + \beta v^{(1)}_n + \beta^2 v^{(2)}_n + \cdots$ with the zeroth-order solution arising in the anticontinuum limit (cf. \cite{fitrakis2007dark,carretero2025nonlinear}).

When $u_+=-0.9$ (with $\beta=0.1$), see Fig.~\ref{fig:neg-values} (B), the initial phase jump imprints different intermediate dynamics than when $u_+=0.9$ (c.f. Fig.~\ref{fig:catalog} (A) for comparison). As shown in the inset (snapshot at $t=100$), a left-propagating coherent structure is generated, and the development of the right-propagating DSW pattern is significantly influenced by large-scale unsteadiness near the central sites. The amplitude of these quasi-periodic oscillations is seen to shrink as $u_+$ is increased, however. As before (when $u_+>0$), at the threshold $u_+=-0.6$, a composite right propagating wave pattern consisting of a DSW-traveling front is observed (c.f. Fig.~\ref{fig:Threshold} for a characterization). The effects of two-phase resonance are again seen to occur at the threshold of around $|u_+|\approx 0.59$ (c.f. inset in Fig.~\ref{fig:neg-values} (D)), leading to the breakdown of the DSW in the composite wave pattern. 

For $u_+\gtrapprox-0.57$ (Figs.~\ref{fig:neg-values} (E), (F)), the unsteady heteroclinic pattern is seen to emerge near the central sites, which represents an earlier parametric transition to this state (than when $u_+>0$). Thus, the heteroclinic transition appears to be dynamically ``favored" when there is an initial phase jump across the central sites. To gain some insight, we display the phase variation across the central sites for a representative dam break with $u_+=-0.5$ in Fig.~\ref{fig:neg_phase_jump_heteroclinic}. Evidently, the polarity of the initial phase jump across the central sites appears to be preserved due to the formation of the heteroclinic transition, in contrast to the case when $u_+>0$, where the initial phase jump is zero.

{The space-time dynamics of this heteroclinic feature exhibits a dominant breathing frequency, accompanied by a rapid decay in oscillation amplitude away from the central site. This phenomenology is illustrated in Fig.~\ref{fig:1_0n1}, which shows the waveform that emerges following the relaxation of the dam break evolution with parameters $u_-=1$, $u_+=-0.1$, and $\beta=0.1$.}

\begin{figure}
    \centering
    \includegraphics[width=\linewidth]{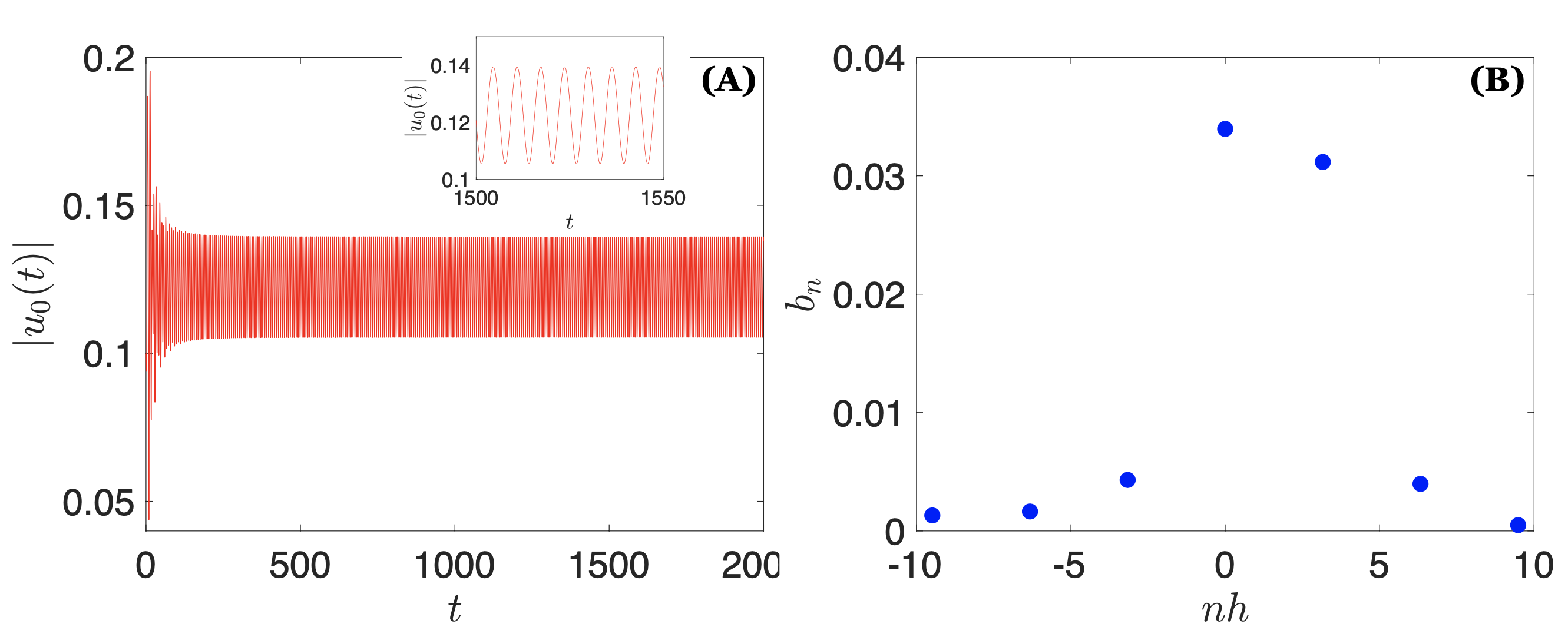}
    \caption{{The heteroclinic, breathing waveform generated from the long-time dynamics of the dam break with $u_-=1$, $u_+=-0.1$ and $\beta=0.1$. (A)the long-time evolution of the central site shows relaxation to a periodic, nearly sinusoidal oscillation with period $T\approx6.4$ and amplitude $b_0\approx 0.034$, (B) The surrounding sites oscillate at nearly the same frequency after relaxation into the breathing state, with oscillation amplitudes decaying away from the central site. }}
    \label{fig:1_0n1}
\end{figure}

\subsection{Vacuum dam break problem}
\label{VACUUM-DAM-BREAK}

In this section, we study the wave patterns emerging from the one-parameter family (in $\beta$) of dam break problems defined by setting $u_+=0$ and $u_-=1$ in Eq.~\eqref{InitialConditionForDNLS}. We study the two extreme limits corresponding to $\beta\gg 1$ and $\beta\ll 1$ in some detail and also examine the transitional wave patterns through numerical simulations.
\subsubsection{Rarefaction waves}
Rarefaction waves are emitted by the vacuum dam break problems provided $\beta>1/(2E^2(\pi/4,2))$, as we saw in Sec.~\ref{Simple-wave-DSW} (c.f. Fig.~\ref{fig:DNLS_out_of_continuum}). In this regime, we therefore agree, at least qualitatively with the continuum NLS case. 
Next, we examine the other extreme regime, namely that
of $\beta \ll 1$ of proximity to the anti-continuum limit, where the dynamical nucleation of the DNLS kink (without a continuum analogue) is recorded.
\subsubsection{Kinks}
We record the emergence of kinks from the Riemann problems, particularly for very small coupling strengths ($\beta\ll1$).  These stationary, real-valued (upto a complex phase factor) states $u_n(t)=v_n \exp(-i\mu t)$ are governed by the difference equation
\begin{equation}
  \textbf{L}_0(v_n)\equiv\mu v_n+\beta(v_{n+1}+v_{n-1}-2v_n)- v_n^3=0.
\end{equation}
\\
In the anti-continuum limit ($\beta=0$) which is often leveraged
to provide an understanding of DNLS dynamical 
states~\cite{kev09}, such  kinks
constitute an exact solution of the form
\begin{align}
    &v_n^{(0)} = \sqrt{\mu}, \;n\leq -1\\\nonumber
    &v_n^{(0)} = 0,\; n\geq 0.
    \end{align}
    Thus, these kinks (at least in the anti-continuum limit) represent a one-parameter family of solutions parametrized by the left-state ($\mu=u_{-}^2$). We compute these using a Newton scheme for finite coupling strength (see, e.g.,~\cite{carretero2025nonlinear}
    for relevant computational details).

Having studied the existence (numerically) of kink solutions, we characterize their stability properties. To do so, we linearize about the kink waveform:
\begin{equation}
    u_n=\exp(-i\mu t)\left[v_n+p_n e^{\lambda t}+q_n^{*}e^{\lambda^* t}\right],
\end{equation}
with the implicit assumption that $|p_n|,|q_n|\ll 1$ 
in some appropriate discrete norm. The linear stability problem then takes the block form
\begin{align} \label{LinearStabilityMatrix}
    \begin{bmatrix}
        i\mu+i\beta \tilde{D}-2i v_n^2 && -i v_n^2\\
        iv_n^2 && -i\mu -i\beta\tilde{D}+2iv_n^2
    \end{bmatrix}\begin{bmatrix}
        p_n\\
       q_n
    \end{bmatrix}=\lambda \begin{bmatrix}
        p_n\\
       q_n
    \end{bmatrix},
\end{align}
The system above satisfies a quartet symmetry, i.e., eigenvalues occur in sets of $\{\lambda,\lambda^{*},-\lambda,-\lambda^{*}\}$ \cite{fitrakis2007dark}.
The eigenvalue problem also satisfies the symmetry $v_n\rightarrow A v_n$, $\beta\rightarrow A^2 \beta$ and $\lambda\rightarrow A^2\lambda$. This symmetry of the linear stability problem suggests the scaling between the critical $\beta$ and $u_-^2$ ($\mu$) as $u_-^2\propto {\beta}$ ($\mu\propto{\beta}$).

We supplement our numerical studies with some analytical approximations in the small coupling limit ($\beta\ll1$). At first, for finite intersite coupling $\beta \ll 1$ akin to \cite{fitrakis2007dark}, we assume a series solution of the form
\begin{equation}
    v_n\sim v_n^{(0)}+\beta v_n^{(1)}+\beta^2v_n^{(2)}.
\end{equation}
Substituting this, we obtain the correction to the solution at $\mathcal{O}(\beta)$
\begin{equation}
    v_0^{(1)}=-\frac{1}{\sqrt{\mu}},\;\;v_{-1}^{(1)}=-\frac{1}{2\sqrt{\mu}};
\end{equation}
The second-order correction is given by
\begin{equation}
    v^{(2)}_n = \begin{cases}
        -\frac{1}{4 \mu^{3/2}}, & n = -2 \\
        -\frac{3}{8 \mu^{3/2}}, & n = -1 \\
        -\frac{3}{2 \mu^{3/2}}, & n = 0 \\
        \frac{1}{\mu^{3/2}}, & n = 1.
    \end{cases}
\end{equation}
which then provides an approximation to the solution $v_{-2}\sim \sqrt{\mu}-\frac{\beta^2}{4\mu^{3/2}}$, $v_{-1}\sim \sqrt{\mu}-\frac{\beta}{2\sqrt{\mu}}-\frac{3\beta^2}{8\mu^{3/2}}$, $v_0\sim -\frac{\beta}{\sqrt{\mu}}-\frac{3\beta^2}{2 \mu^{3/2}}$ and $v_{1}\sim \frac{\beta^2}{\mu^{3/2}}$. 

Next, we turn to the leading order corrections to the stability spectrum near the anti-continuum limit. In the anti-continuum limit ($\beta=0$), the eigenvalues of the stability matrix (Eq.~\eqref{LinearStabilityMatrix}) can be obtained (explicitly) either by looking at the zero and nonzero sites, which are associated with $\lambda=\pm i\mu $ and $\lambda=0$, respectively. Now, in the finite (but small) $\beta$ case, an approximation to the spectrum can be obtained. At first, we note the emergence of the continuous spectrum given by $-i\sqrt{16 \beta^2+8\beta u_-^2}\leq \lambda \leq  i\sqrt{16 \beta^2+8\beta u_-^2}$. Furthermore, we consider the perturbations of the eigenvalue $\pm i\mu$ as we depart gradually from the anti-continuum limit. For this, we consider the $4\times 4$ diagonal block associated with the sites $n=-1,0$.
\begin{equation}
    \tilde{E} =i\begin{bmatrix}
        -\mu & \beta-\mu & \beta & 0 \\
        \mu-\beta & \mu & 0 & -\beta\\
        \beta & 0 &\mu-2\beta & 0 \\
        0 & -\beta & 0 & -\mu+2\beta\\
    \end{bmatrix}
\end{equation}
Expressions for the eigenvalues of $\tilde{E}$ are given by $\lambda=i\Lambda$, where $\Lambda^2$ is defined below
\small{
\begin{align}
    \label{EIG-VALUES}&\Lambda_{1,2,3,4}^2\\\nonumber&=\frac{{5\beta^2\pm\sqrt{37\beta^4-52\beta^3\mu+42\beta^2\mu^2-12\beta \mu^3+\mu^4}-2\beta\mu+\mu^2}}{2},
\end{align}}
where $|\Lambda_{1,2}|^2>|\Lambda_{3,4}|^2$. {For sufficiently small values of $\beta$, two of the eigenvalues are embedded within the continuous spectrum. However, we note that the other pair of eigenvalues lies outside the continuous spectrum band.} 
The {latter} pair of eigenvalues outside collide with the continuous spectrum band at $\tilde{\beta}_c\approx 0.0771\mu$. Since the colliding spectra have \textit{opposing} Krein signatures, a Hamiltonian-Hopf bifurcation is known to occur \cite{van1990hamiltonian}, producing a quartet of complex eigenvalues. This suggests that kinks are subject to an oscillatory instability. Furthermore, as an additional note, the quantity $37\beta^4-52\beta^3\mu+42\beta^2\mu^2-12\beta \mu^3+\mu^4$ within the square root in Eq.~\eqref{EIG-VALUES} changes sign at a larger $\beta\approx0.145\mu$, while staying positive at smaller values of $\beta$.
 One can also utilize the scaling symmetry of the eigenvalue problem ($v_n\rightarrow A v_n$, $\beta\rightarrow A^2 \beta$ and $\lambda\rightarrow A^2\lambda$) to find the aforementioned critical threshold. In particular, we may assume $\mu=\alpha \beta$, where $\alpha$ is an appropriate scaling coefficient that is to be determined.

{For completeness, we also derived the higher-order correction to the spectrum of the stability operator in \eqref{LinearStabilityMatrix} by analyzing contributions from the sites $n=-2,-1,0,1$ , retaining terms up to $\mathcal{O}(\beta^2)$. This refinement yields a slightly more accurate prediction for the onset of the Hamiltonian-Hopf bifurcation, at $\tilde{\beta}_c\approx 0.0684 \mu$. Besides, these calculation predict the presence of four embedded eigenvalues within the continuous spectrum when $\beta<\tilde{\beta}_c$. i.e., 
presumably in extending the calculation to more sites,
in addition to the eigenvalue pair of interest, we are
progressively accounting for modes within
the continuous spectrum.} {A sharp estimate of the (in)stability threshold can be obtained by tracking the collision of the continuous spectral bands associated with the zero background ($(1-4\beta)\leq |\lambda|\leq 1$) and the non-zero background ($|\lambda|\leq \sqrt{16\beta^2+8\beta u_{-}^2}$) of the kink, which carry opposite Krein signatures. For $u_-=1$, this is seen to occur for $\tilde{\beta}_c=1/16= 0.0625$, determined by $\sqrt{16\beta^2+8\beta}=1-4\beta$. Checking this against the numerically predicted threshold (i.e. examining the parametric ($\beta$) variation of eigenvalues of Eq.~\eqref{LinearStabilityMatrix}), we find that for $N=800$ this transition to complex quartets with associated real part $\sim 10^{-6}$ occurs at $\beta\approx 0.0629$. }

An extensive stability analysis suggests a dependence of the stability spectra on the number of sites present in the lattice, for $\beta>\tilde{\beta_c}$ (analogously to the case of dark solitons \cite{johansson1999discreteness}). However, for $\beta$ close to the threshold $\beta_c$ the instability growth rates are minimal. For example, with $1200$ sites, the kink characterized by $\beta=0.1$ (and $\mu =1$), has ${\rm max}({\Re(\lambda)})\approx 1.22\times 10^{-4}$. Thus, the kinks are observed resulting from the Riemann problems for reasonably long times ($t=1000$) in Fig.~\ref{fig:catalog} (F). However, for Riemann problems with increasing $\beta$ these structures are subject to progressively more pronounced instability growth rates. This can be leveraged to explain why they are not observed to persist even for short times therein (c.f. Fig.~\ref{fig:instability}).

{To verify the linear stability predictions associated with the unstable kinks, we seeded their profiles with the maximally unstable eigenmode (i.e. $u_n=v_n+\epsilon(p_n+q_n^{*})$, $\epsilon \ll 1$) and tracked their complex-valued growth rate numerically by monitoring  ${\rm max}_n||u_n(t)|-v_n|$. To corroborate this against the analytical result, we perform a numerical fit (for $\epsilon$) in the dynamical linear stability prediction $u_n=\exp(-i\mu t)[v_n+\epsilon(p_n e^{\lambda t}+q_n^*e^{\lambda^* t} )]$ where $\lambda$ is the eigenvalue associated with the maximally unstable eigenmode. Note that not only do we capture the exponential growth (in time) ${\rm Re}(\lambda)\approx 0.0302$, but also capture the oscillatory part of the instability ${\rm Im}(\lambda)\approx 0.258$ accurately (see also Fig.~\ref{fig:plot_growth_rate} for the case $\beta=0.18$).  }
\begin{figure}
    \centering
    \includegraphics[width=\linewidth]{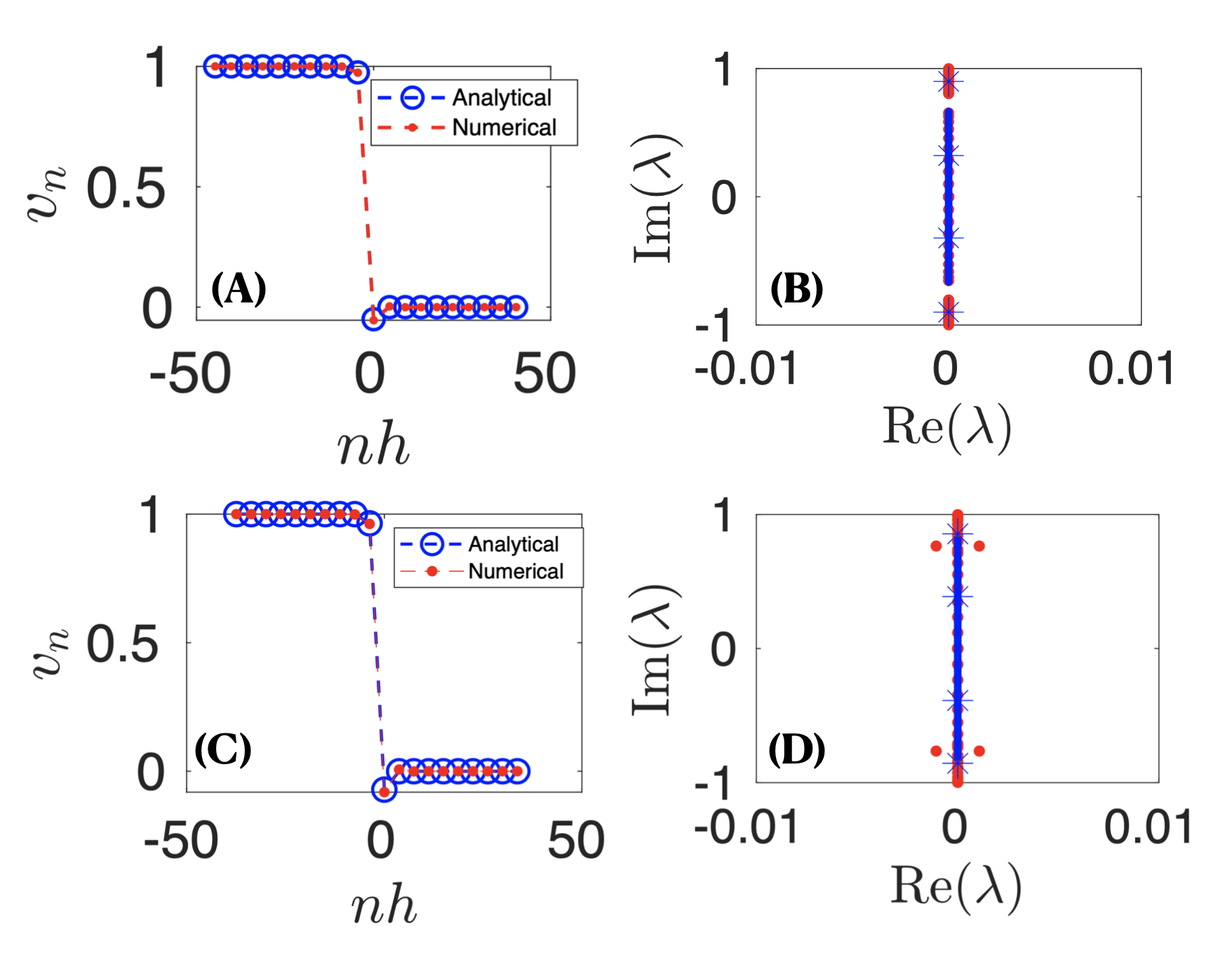}
    \caption{(A) Kink computed numerically (in red) and found analytically in the small $\beta$ limit (in blue), for $\beta=1/20$, (B) The numerically obtained spectrum (in red) compared to the analytical formulas {[perturbation theory result]} for the same $\beta=1/20$ (discrete eigenvalues: blue stars, continuous spectrum: blue dots), 
 (C) Kink computed numerically (in red) and found analytically in the small $\beta$ limit (in blue), for $\beta=1/14$, which lies in the regime {across} the Hopf bifurcation point, (D) The eigenspectrum (analytical: blue versus numerical: red) shown for the same $\beta=1/14$. }
    \label{fig:Kink}
\end{figure}
\begin{figure}
    \centering
    \includegraphics[width=\linewidth]{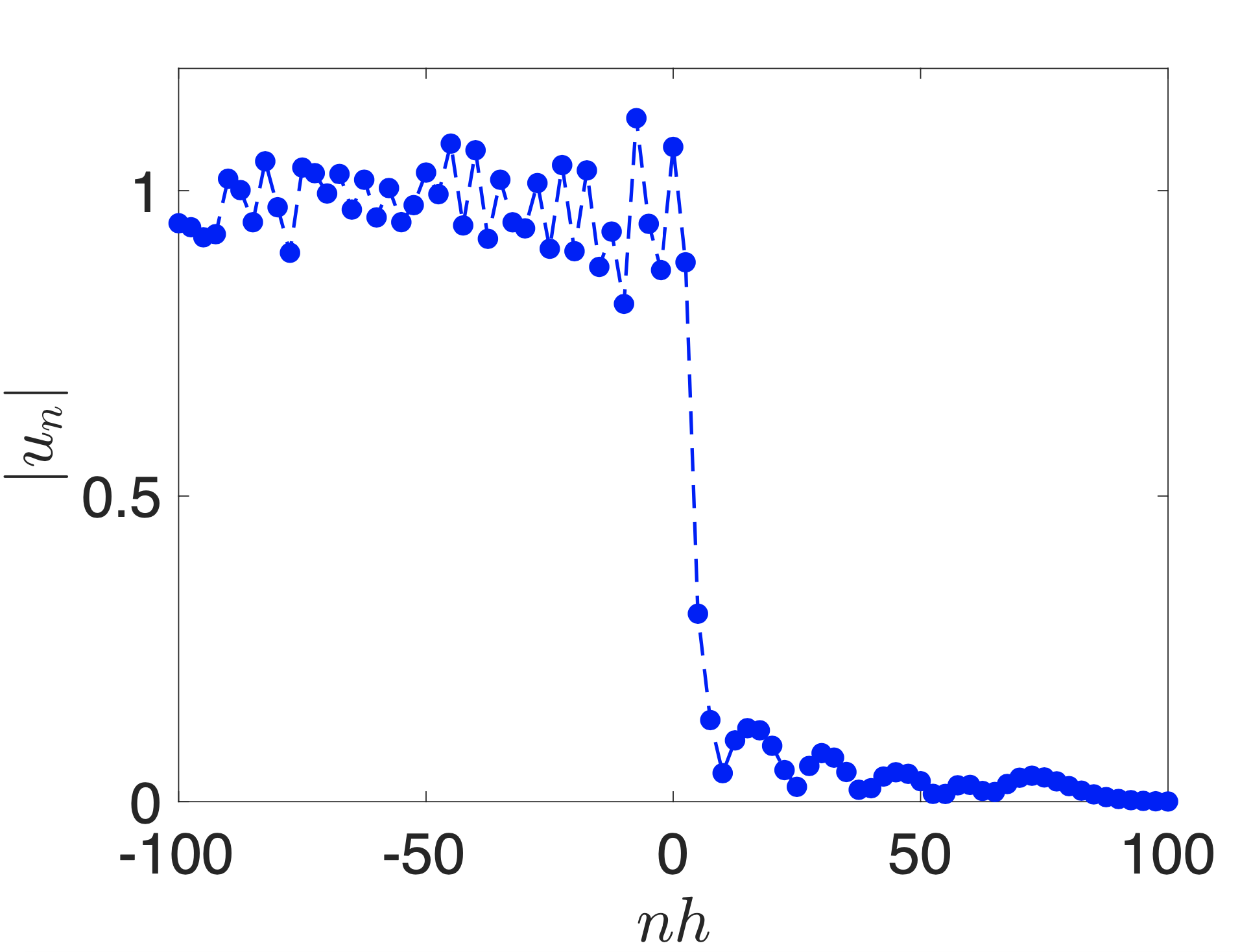}
    \caption{A snapshot of the dynamics at $t=100$ resulting from the dam break with $\beta=0.16$. Here, the dynamics do not settle into the kink state, perhaps due to the prominent instability growth rates associated with the stationary kink.}
    \label{fig:instability}
\end{figure}

\subsubsection{Vacuum dam breaks for intermediate $\beta$}
In prior subsections, we saw that for large coupling strength ($\beta>1/(2E^2(\pi/4,2))$) dam breaks characterized by $u_-=1$ develop into rarefaction waves, qualitatively similar to the continuum defocusing NLS. 
However, for small coupling strengths, dam breaks were seen to develop into stationary kink states around the central sites. Furthermore, the transition from the kink states to richer, unsteady dynamics as a result of increasing $\beta$ (Fig.~\ref{fig:instability}) was not as sharp.

Next, we probe the parameter space of vacuum dam breaks for $\beta>0.2$. We display a catalog of such wave patterns at $t=1000$ in Fig.~\ref{fig:vacuum-dam}.
A representative dynamical scenario corresponds to a composite rarefaction-traveling DSW wave pattern that is overwhelmed by large-scale instabilities (Fig.~\ref{fig:vacuum-dam} (B)). Further increasing the coupling strength reveals a coherent, composite rarefaction-traveling DSW pattern (Fig.~\ref{fig:vacuum-dam} (C)). In particular, the inset reveals the structure of the traveling DSW, whose existence (numerical) and stability characterizations require the development of further techniques
in future studies. 

\begin{figure}
    \centering
    \includegraphics[width=\linewidth]{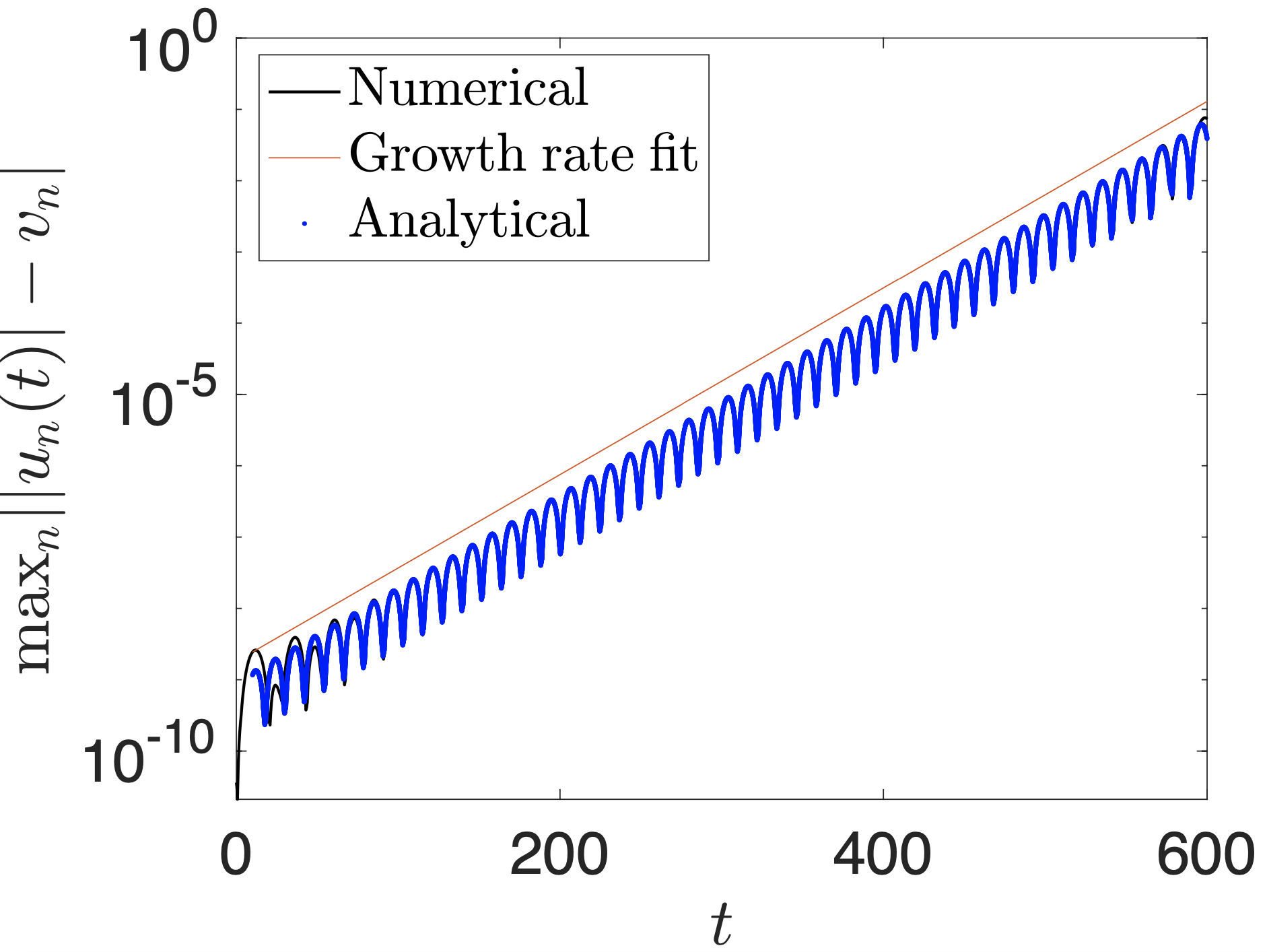}
    \caption{{The growth of the (oscillatory) instability mode associated with a kink characterized by $\beta=0.18$ and $N=800$ (number of sites). The extracted growth rate (numerically, see red solid line) agrees with the analytical prediction $\underset{\lambda_i}{\rm max}[{\rm Re}(\lambda)]\approx 0.0302$ to within $0.2\%$. Likewise, the imaginary part of the same eigenvalue ($\approx 0.258$) accurately captures the oscillatory nature of the unstable mode, exhibiting the same level of relative accuracy, as evidenced by the agreement between the (dynamical) quasi-analytical (blue dots) and numerical (black solid) curves in the time interval $t\in [100,600)$. }}
    \label{fig:plot_growth_rate}
\end{figure}

\begin{figure}
    \centering
    \includegraphics[width=\linewidth]{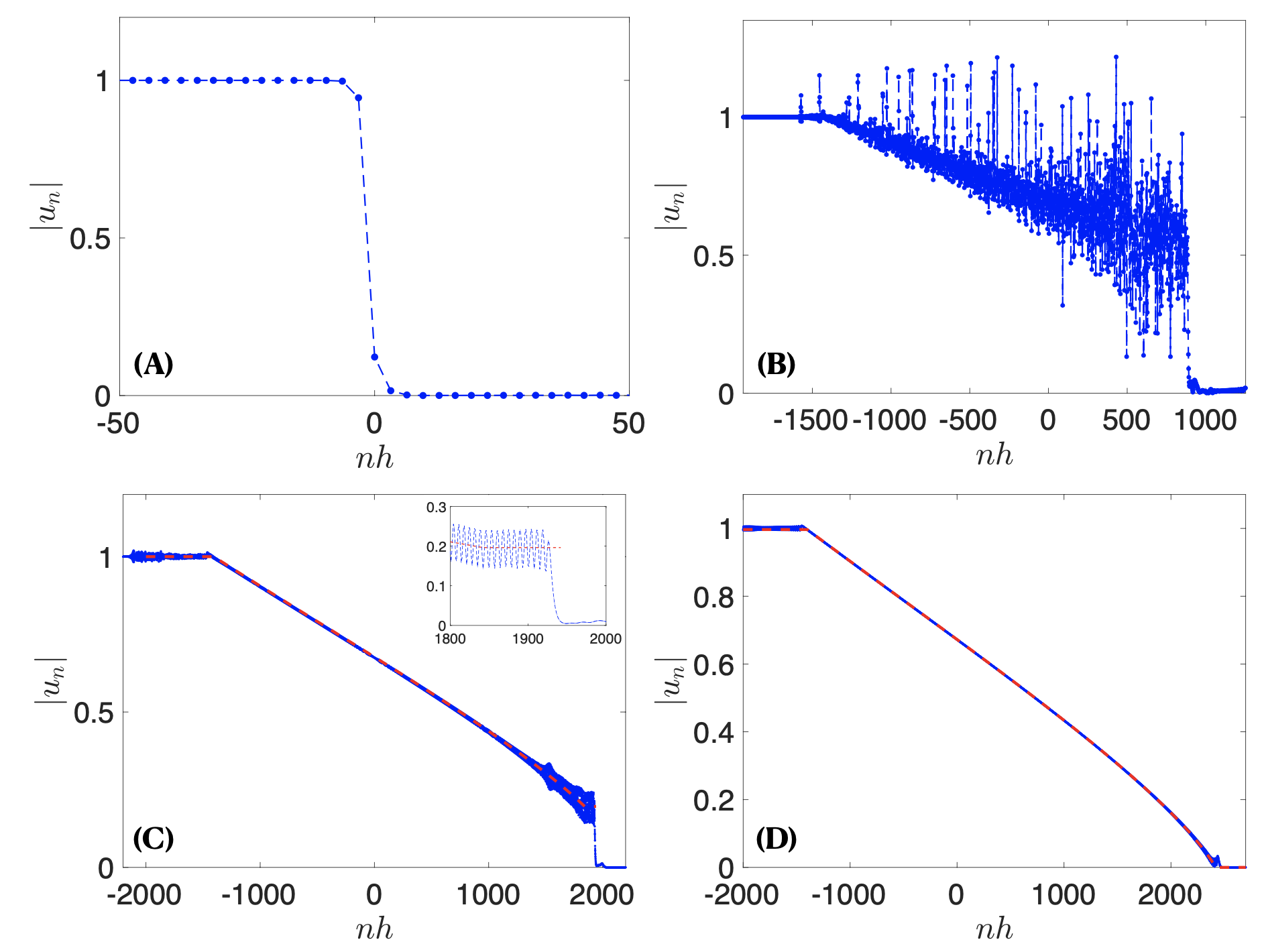}
    \caption{Snapshots (at $t=1000$) of features emitted from the vacuum dam break ($u_+=0$), as $\beta$ is varied. (A) For $\beta=0.1\ll 1$, we observe stationary discrete NLS kinks, (B) For $\beta=0.4$, we observe a composite wave pattern overwhelmed by large-scale instability effects, (C) For $\beta=1$, we observe a composite rarefaction-travelling DSW structure. The inset displays the internal variation of the traveling DSW, (D) A rarefaction wave is observed for $\beta=1.5>1/(2E^2(\pi/4,2))$. The simple wave solutions of the one-phase modulation equations (Eqs.~\eqref{One-phase-modn}) are overlaid in red-dashed line for comparison.  }
    \label{fig:vacuum-dam}
\end{figure}

 \section{Conclusions and scope for future work}
 
In this work, we have explored the nucleation of DSWs in the discrete nonlinear Schr{\"o}dinger equation Eq.~\eqref{EquationOfMotionForDNLS}, with a view towards generating an overarching phenomenological map of 
the rich nonlinear wave features of the model, in distinct parameter regimes. To do so, we have examined a two-parameter family of \textit{dam-break} problems characterized by the coupling strength $\beta$ and right hydrodynamic background $u_+$, while holding $u_-=1$. The study of shock waves not only in the DNLS, but also in the closely related {Salerno} model has also received attention in the earlier works of \cite{konotop1997dark,konotop2000shock,kamchatnov_dissipationless_2004}, wherein, asymptotic models were deployed to shed light on the types of shock structures and time for shock formation in different parameter regimes. However, many of these early studies were focused on the 
vicinity of the continuum limit and did not explore the highly
discrete anti-continuum realm where much of the rich phenomenology
uncovered herein was identified.
In spirit, our work is close to the notable work of~\cite{panayotaros2016shelf} which identified some novel wave patterns arising in the DNLS framework that do not have a continuum counterpart. In our present work, we extend the current state-of-the-art understanding of DSW formation in the DNLS by leveraging recent developments on the Whitham modulation equations for generalized NLS models~\cite{hoefer_shock_2014,sprenger2024whitham}. 
We combine this analysis with systematic numerical computations 
spanning the different regimes of our two-parameter space
This approach has enabled us to unveil
novel wave structures such as traveling DSWs, compound and composite waves (rarefaction-traveling DSW, etc.) and, patterns (such as those in Figs.~\ref{fig:catalog} (E), \ref{fig:neg-values} (E)) reminiscent of unsteady \textit{Whitham} shocks \cite{sprenger2024hydrodynamics}. There is a phase jump across such heteroclinic transitions (see, for instance, Fig.~\ref{fig:neg_phase_jump_heteroclinic}), and thus, these are observed to arise more naturally from the dam breaks characterized by $u_+<0$. A careful characterization of their existence through numerical means and a study of their stability is important for future work. The zero-amplitude limit  $u_+=0$ of these unsteady features is the stationary kink. These waveforms suffer from instability due to a Hamiltonian-Hopf bifurcation for sufficiently large $\tilde{\beta}_c$, and are thus never seen to emerge from the dam breaks with large $\beta\gg \tilde{\beta}_c$. In this work, we have also observed the so-called ``two-phase" modulational instabilities \cite{sprenger2024whitham} arise in the context of Riemann problem dynamics. In particular, {the suggested} \textit{two-phase} resonance, was seen to play a crucial role in the breakdown of DSW for $\beta\ll \beta_c$. {An important direction for future work is a detailed study of the short-time development of instabilities due to two-phase resonance.} In such scenarios, the nonlinear stage of the generalized MI was seen to  nucleate weakly pulsating breather trains, whose small amplitude limit appeared to be approximated by solitary waves of weakly nonlinear reductions. 
Large amplitude variants of these states were also identified and
were found to possess  a surprising and distinctive ``box" shape (c.f. Fig.~\ref{fig:DNLS_center_of_mass_in_box_type_breathers}). Crucially, the weak pulsations of these nearly traveling wave-like features 
render them a particularly challenging structure to identify in an
``exact'' form, as methods for traveling waves would need
to be combined with ones producing DNLS 
breathers, as, e.g., in the work of~\cite{johansson1997existence}
or the recent one of~\cite{LYTLE2025103547}.


{Besides being a spatially discretized representation of the defocusing (repulsive) NLS, the discrete NLS is also physically relevant in its own right. In particular, it is an envelope approximation model for light propagating through a waveguide while coupled to its nearest neighboring waveguides (tight binding approximation) with the \textit{self-phase modulation} dominating the \textit{cross-phase modulation} \cite{kev09}. Such a tight-binding approximation may also render the Gross-Pitaevskii (GP) equation with a periodic potential into a discrete NLS-type model. Notably, such a GP equation may describe the dynamics of the superfluid trapped by an additional optical lattice \cite{kevrekidis2003stability}. 
Indeed, current developments have enabled the accurate and time-resolved
visualization of discrete solitary waves in such optical lattice
atomic settings, e.g., in the very recent work of~\cite{haller}.
A direction under ongoing investigation corresponds to generalized Riemann problems in superfluids trapped by an additional optical lattice. }

In addition to the experimental applicability of the relevant model
which begs the question of revisiting relevant experiments 
(considered, e.g., in~\cite{jia_dispersive_2007}), our work
raises a number of further questions for future study.
Indeed, many among the structures we examined merit further
exploration. These encompass, but are not limited to, traveling
DSWs, composite DSWs, more complex (potentially higher
genus) dynamical states, heteroclinic transitions featuring breathing
behavior, dynamically moving (appearing as nearly traveling
states) of different shapes and forms, including some rather
exotic ones such as the box-shaped state. Remarkably, all of these
states were found to spontaneously arise in the numerical experiments
with Riemann initial data considered herein. We believe that the
present investigation may spearhead the further study of associated
states and be a springboard for further developments in the 
emerging field of discrete dispersive hydrodynamics.
 \section{Acknowledgements}
 We would like to thank Dr. Patrick Sprenger and Prof. Mark A. Hoefer for the insightful discussions. 
\bibliographystyle{apsrev4-2}
\bibliography{Refs_clean}
\end{document}